\documentclass[natbib,twocolumn]{aastex61}

\newcommand{\Ppuls}{$P_{\rm{p}}$}
\newcommand{\mdash}{\hbox{---}}
\newcommand{\Msol}{\mathcal{M}_\odot}

\received{December 1, 2017}
\revised{April 20, 2018}
\accepted{May 15, 2018}
\submitjournal{ApJ}

\shorttitle{Cepheid Distance Scale Bias due to Blending}
\shortauthors{R.I.~Anderson \& A.G.~Riess}

\begin{document}

\title{On Cepheid Distance Scale Bias due to Stellar Companions and Cluster Populations}

\correspondingauthor{Richard~I. Anderson}
\email{randerso@eso.org}

\author[0000-0001-8089-4419]{Richard~I. Anderson}
\affiliation{European Southern Observatory, Karl-Schwarzschild-Str. 2, D-85748 Garching b. M\"unchen, Germany}
\affiliation{Department of Physics and Astronomy, The Johns Hopkins University, 3400 N Charles St, Baltimore, MD 21218, USA}

\author{Adam~G. Riess}
\affiliation{Department of Physics and Astronomy, The Johns Hopkins University, 3400 N Charles St, Baltimore, MD 21218, USA}
\affiliation{Space Telescope Science Institute, 3700 San Martin Dr, Baltimore, MD 21218, USA}

\begin{abstract}

State-of-the art photometric measurements of extragalactic Cepheids account for the mean additional light due to chance superposition of Cepheids on crowded backgrounds through the use of artificial star measurements.  However, light from stars physically associated with Cepheids may bias relative distance measurements if the changing spatial resolution along the distance ladder significantly alters the amount of associated blending. We have identified two regimes where this phenomenon may occur: Cepheids in wide binaries and open clusters.

We estimate stellar association bias using the photometric passbands and reddening-free Wesenheit magnitudes used to set up the distance scale. For wide binaries, we rely on Geneva stellar evolution models in conjunction with detailed statistics on intermediate-mass binary stars. For the impact of cluster stars, we have compiled information on the clustered Cepheid fraction and measured the typical cluster contribution in M31 via deep {\it HST} imaging provided by the PHAT project.

We find that the dominant effect on the distance scale comes from Cepheids in clusters, despite cluster Cepheids being a relatively rare phenomenon. Wide binaries have a negligible effect of $0.004\%$ on $H_0$ for long-period Cepheids observed in the near-infrared or when considering Wesenheit magnitudes. We estimate that blending due to cluster populations has previously resulted in a $0.23\%$ overestimate of $H_0$. Correcting for this bias, we obtain $H_0 = 73.07 \pm 1.76\, \rm{km\,s^{-1}\,Mpc^{-1}}$, which remains in $3.3\sigma$ tension with the {\it Planck} value. We conclude that stellar association bias does not constitute a limit for  measuring $H_0$ with an accuracy of $1\%$.

\end{abstract}


\section{Introduction} \label{sec:intro}

The extragalactic distance scale provides a crucial measure of the
present-day expansion rate of the universe, i.e, the Hubble constant $H_0$.
Historically, there has been great interest in measuring $H_0$ with ever
improving accuracy \citep[for recent reviews, cf.][]{2010ARA&A..48..673F,2013PhT....66j..41L}. A significant improvement was 
the \textit{Hubble Key Project} \citep{2001ApJ...553...47F}, which reached the $10\%$ accuracy level. Since then, improved observing strategies that favor data homogeneity, better error analysis and propagation, larger samples of Cepheids in SN-host galaxies \citep[henceforth: R+11]{2011ApJ...730..119R}, and better trigonometric parallaxes of Cepheids in the Milky Way have enabled considerable improvements, so that the current accuracy of $H_0$ now
figures at $2.4\%$ \citep[$73.24 \pm 1.74\,\rm{km\,s^{-1}\,Mpc^{-1}}$, cf.][henceforth: R+16]{2016ApJ...826...56R}. In the R+16 implementation, classical Cepheids provide a geometric calibration for type-Ia supernovae peak luminosities and covariance is taken into account by modeling the distance scale globally. 
A distance-scale independent estimate of $H_0$ with $3.8\%$ precision ($72.8 \pm 2.4\,\rm{km\,s^{-1}\,Mpc^{-1}}$) has been obtained using time delays observed in gravitationally lensed quasars using a prior on $\Omega_M$ \citep{2017MNRAS.465.4914B}, which is in good agreement with R+16. Exchanging SNIa optical magnitudes with near-IR data also yields a result consistent with this value of $H_0$ \citep[$72.8 \pm 1.6 \pm 2.7\,\rm{km\,s^{-1}\,Mpc^{-1}}$]{2018A&A...609A..72D}.
Finally, gravitational wave events may provide single digit accuracy on $H_0$ in the future \citep{2012PhRvD..86d3011D,2017Natur.551...85A}. Of course, many complementary efforts are under way to accurately measure $H_0$, e.g. using standard candles belonging to older stellar populations \citep[e.g.][]{2016ApJ...832..210B}.

There are two key motivations for continuing to push $H_0$ accuracy. First, $H_0$ can serve as a powerful prior for analyses of the Cosmic Microwave Background (CMB) and knowing $H_0$ to $1\%$ accuracy would significantly improve the uncertainties on the dark energy equation of state $\sigma_w$ \citep{2012arXiv1202.4459S,2013PhR...530...87W,2016PhRvD..93f3009M}, which is crucial for understanding the origin and nature of the universe's accelerated expansion \citep{1998AJ....116.1009R,1999ApJ...517..565P}. 

Second, R+16 have shown that $H_0$ measured directly using a state-of-the-art distance scale now differs by $3.4\sigma$ from the value inferred from the {\it Planck} satellite's measurements of the CMB assuming $\Lambda$CDM \citep{2016A&A...594A..13P,2016A&A...596A.107P}. This has since been confirmed independently \citep{2018MNRAS.476.3861F,2017JCAP...03..056C,2018MNRAS.tmp..707F}. Moreover, this difference cannot be explained solely by invoking systematic errors in the {\it Planck} data \citep{2018ApJ...853..119A}. If this difference is amplified by even more accurate determinations of $H_0$, then such a discrepancy could lead to the exciting conclusion that the presently-accepted cosmological model is incomplete. 

Of course, detailed investigations of systematics intervening in $H_0$ measurements are required before new physics may be credibly invoked to explain this observed difference. Large efforts are already underway to this end, focusing on all aspects of the distance ladder. For Cepheids specifically, possible non-linearities or metallicity-dependence of the Leavitt Law, i.e., the Period-luminosity Relation of Cepheids \citep[PLR]{1912HarCi.173....1L}, have received much recent attention \citep[e.g.][]{2004A&A...424...43S,2004ApJ...608...42S,2004A&A...415..531S,2006ApJ...650..180N,2013MNRAS.431.2278G,2013ApJ...764...84I,2015ApJ...799..144K,2016MNRAS.457.1644B,2017ApJ...842..116W}. Yet, none of these effects have been confirmed to significantly impact $H_0$.
  
\begin{table*}
\begin{tabular}{@{}lrll@{}}
\hline
\hline
Definitions \\
\hline
 & $M_{\rm{MW}}(P_{\rm p})$ & $= m_{\rm{Cep,MW}} - \mu_{0,\rm{Gaia}}$ & $= - 2.5\cdot \log{(\mathcal{F}_{\rm{Cep}}(P_{\rm p}) + \mathcal{F}_{\rm{Comp,near}})} - 5\log{d_{\rm{Cep}}} + 5$ \\
 & $M_{\rm{LMC}}(P_{\rm p})$ & $= m_{\rm{Cep,LMC}} - \mu_{0,\rm{LMC}}$ & $= - 2.5\cdot \log{(\mathcal{F}_{\rm{Cep}}(P_{\rm p}) + \mathcal{F}_{\rm{Comp,near}} + \mathcal{F}_{\rm{Comp,wide}})} - 5\log{d_{\rm{LMC}}} + 5$ \\
 & $M_{\rm{SN}}(P_{\rm p})$ & $= m_{\rm{Cep,SN}} - \mu_{0,\rm{Gaia}}$ & $= - 2.5\cdot \log{(\mathcal{F}_{\rm{Cep}}(P_{\rm p}) + \mathcal{F}_{\rm{Comp,near}} + \mathcal{F}_{\rm{Comp,wide}} + \mathcal{F}_{\rm{Clusters}})}\ -$ \\
 & & & $\ \ \ - m_{\rm{Cep,MW}} + M_{\rm{MW}}(P_{\rm p})$ \\
 \hline
 & $\Delta M_{\rm{SN}}$ & $= M_{\rm{SN}} - M_{\rm{MW}}$ & $= f_{wb} \cdot \Delta M_{\rm{wb}} + f_{CC} \cdot \Delta M_{\rm{Cl}}$ \\
 \hline 
& & $f_{\rm{wb}}$ & fraction of Cepheids with wide ($400 \lesssim a_{\rm{rel}} \lesssim 4000$\,AU) companions, cf. \S\ref{sec:binaryfraction} \\
& & $\Delta M_{\rm{wb}}$ & typical brightening by a companion on a wide orbit, cf. \S\ref{sec:typicalcompanionbias}\\
& & $f_{\rm{CC}}$ & fraction of Cepheids occurring in clusters, cf. \S\ref{sec:CCfrequency} \\
& & $\Delta M_{\rm{Cl}}$ & typical brightening due to cluster stars. cf. \S\ref{sec:M31ClusterCepheidPhotometry}\\
\hline
\hline
 \end{tabular}
\caption{Definition of stellar association bias. $d$ denotes distance, $\mathcal{F}$ flux received, $m$ apparent magnitude, $M$ absolute magnitude, $\mu$ distance modulus, and $P_{\rm p}$ pulsation period. Subscript $_{\rm{SN}}$ refers to a typical SN-host galaxy, e.g. in the {\it SH0ES} project (R+16). $M_{\rm{SN}}$ is assumed to be computed using a Galactic PLR calibrated using {\it Gaia} parallaxes of Milky Way (MW) Cepheids.}
\label{tab:fluxcontributors}
\end{table*}

The crucial geometric footing for the cosmic distance scale is currently being rebuilt thanks to technical advances in accurately measuring trigonometric parallaxes of classical Cepheids. Specifically, observations made in spatial scanning mode using {\it HST/WFC3} \citep[Riess et al. in prep.]{2014ApJ...785..161R,2016ApJ...825...11C} are used to measure parallax of a number of Galactic Cepheids with $30 - 40 \mu$arcsec accuracy \citep[cf. also][]{2016ApJS..226...18A}. On an even larger scale, the ESA mission {\it Gaia} is currently measuring parallax of approximately $300$ Galactic Cepheids with better than $3\%$ accuracy \citep{GaiaMission,2016A&A...595A...2G,2017A&A...605A..79G}. Thus, the Galactic calibration of the Leavitt law will be considerably improved compared to the previous calibration based on $10$ Cepheids with parallaxes known to better than $10\%$ accuracy \citep{2007AJ....133.1810B}. 

The spatial coincidence of multiple light sources within a detector resolution element or PSF (henceforth: {\it blending}) becomes increasingly likely with distance, since the physical scale (in pc) of a single pixel increases with distance for a fixed plate scales (in arcsec per pixel). Thus, apparent Cepheid magnitudes are expected to be increasingly affected by ``parasitic'' flux contributions the farther away the galaxy in which they reside. 

Blending can occur due to chance superposition (e.g. field stars in a distant galaxy of interest) or physical association (e.g. companion stars in binaries or cluster member stars). Notably, blending due to chance superposition can be effectively corrected using local, artificial star tests \citep[cf. Sec.\,2.3 of][]{2009ApJ...699..539R} that estimate the average light contribution per pixel due to field stars near the object of interest. However, stars physically associated with Cepheids are necessarily close (within a few parsecs) and thus potentially unresolved, and the properties of their light contribution may differ from that of the (possibly crowded) field stars. Given a fixed plate scale, the ability to resolve physically associated stars depends on distance, so that flux contributed by physically associated stars cannot always be estimated directly from the observations of each galaxy. Although it would be possible to apply a homogeneous aperture of fixed physical scale to all galaxies (including the MW), this would a) add considerable noise (in particular in the MW) and b) lack the external view of MW Cepheids analogous to their extragalactic counterparts. Table\,\ref{tab:fluxcontributors} provides an overview of the physically associated objects blending into a Cepheid's PSF on different ``rungs'' of the distance ladder.

Several previous studies have targeted blending effects due to chance superposition by estimating the brightening of Cepheids as a function of angular resolution by comparing apparent magnitudes measured using observations from the ground and from Space \citep{1999astro.ph..9346S,2000AJ....120..810M,2001astro.ph..3440M,2005ApJ...634.1020B,2005MNRAS.358..883K,2007A&A...473..847V,2015ApJ...813...31S}. The claim that blending causes significant distance scale bias presented in some of these studies\hbox{---}and refuted by others \citep[e.g.][]{2000ApJ...530L...5G}\hbox{---}requires revision for several reasons \citep[cf. also][]{2000PASP..112..177F}. First, blending by field stars is routinely corrected using artificial star tests. Second, the distinction between physically associated objects, such as cluster members and companion stars, and nearby field stars had previously remained somewhat unclear. This distinction is crucial, however, and can be made much more clearly now thanks to updated statistics on stellar multiplicity\mdash in particular regarding wide companions\mdash and cluster membership of Cepheids, as well as deep {\it HST} imaging of a significant portion of M31. Notably, the common assertion that Cepheids frequently occur in star clusters has remained largely unchecked and has led to the erroneous (cf. Sec.\,\ref{sec:CCfrequency}) interpretation that previous blending estimates were primarily sensitive to physically associated objects instead of chance alignment. Finally, the discussion of blending-related issues has not yet been updated to reflect the current state-of-the-art calibration of the distance scale, which utilizes artificial star corrections, or the typical photometric passbands used to measure $H_0$. Thus, this {\it article} seeks to clarify the impact of physically associated stars on the distance scale.

\begin{table}
\centering
\begin{tabular}{@{}cccccc@{}}
\hline\hline
$\Delta\mathcal{F}$ & $m$ & $M$ & $\mu = m - M$ & $d = 10^{0.2(\mu +5)}$ & $H_0 = v / d$\\
$\uparrow$ & $\downarrow$ & $=$ & $\downarrow$ (closer) &  $\downarrow$ (closer) & $\uparrow$ (faster) \\ \hline\hline
\end{tabular}
\caption{Visualizing the impact of added flux contributions to Cepheids in SN Ia hosts due to wide companion stars or cluster members on $H_0$, in case the Leavitt law is calibrated using Cepheids that are not subject to such contamination.}
\label{tab:biasdirection}
\end{table}

To this end, we estimate the photometric bias due to blending of physically associated objects and its influence on the $H_0$ measurement presented in R+16. The bias exists because some physically associated objects (wide binaries and cluster stars) are spatially resolved on Galactic scales, whereas they cannot be identified or de-biased in distant galaxies that set the luminosity zero-point for type-Ia supernovae. Thus, stellar association bias leads to systematic differences between Leavitt laws observed along the distance ladder. Table\,\ref{tab:biasdirection} illustrates the direction of this bias for different quantities of interest.

We estimate separately the bias contribution due to companion and cluster stars. \S\ref{sec:binaries} details the estimation of bias due to companion stars on wide orbits ($a_{\rm{rel}} > 400$\,au) using binary statistics and state-of-the-art stellar evolution models. We adopt a synthetic approach for wide binaries, since photometric {\it HST} observations of Galactic Cepheids are scant, whereas binary statistics of intermediate-mass stars have been studied in detail. Moreover, most {\it visual} companions to Cepheids are likely not physically associated \citep{2016AJ....151..108E,2016AJ....151..129E}. \S\ref{sec:ClusterCepheids} presents our empirical estimation of bias due to cluster stars. To this end, we review the occurrence rate of Cepheids in clusters in the Galaxy (MW), Large Magellanic Cloud (LMC), Small Magellanic Cloud (SMC), and the Andromeda galaxy (M31), and employ deep {\it HST} imaging of M31 by the {\it PHAT} project \citep{2012ApJS..200...18D} to obtain an empirical estimate of the average light contribution from Cepheid host cluster populations. Unless otherwise specified, the term `Cepheid' refers to type-I (classical) Cepheids pulsating in the fundamental mode throughout this paper.
\S\ref{sec:discussion} discusses relevant uncertainties and limitations of this work and provides recommendations for mitigating stellar association bias. \S\ref{sec:summary} summarizes our results.

\section{Binary Stars}
\label{sec:binaries}

Cepheids frequently occur in binary or higher order multiple systems \citep[e.g.][]{2003ASPC..298..237S,2015A&A...574A...2N}, and despite long-standing efforts to detect companions even well-studied cases such as the prototype $\delta$\,Cephei occasionally hold surprises \citep{2015ApJ...804..144A}. Most companions of Galactic Cepheids are not spatially resolved, although several companions have been directly detected using long baseline optical interferometry \citep{2015A&A...579A..68G}. Conversely, most {\it visual} companions\mdash i.e., spatially resolved stars located near a Cepheids\mdash appear to be not physically associated, since physically bound companions seem to be limited to relative semimajor axes $a_{\rm{rel}} \lesssim 4000$\,au \citep{2016AJ....151..108E,2016AJ....151..129E}. 

Companions of high-interest long-period Galactic Cepheids \citep[typical distance $\sim 2.5$\,kpc, cf.][]{2014ApJ...785..161R,2016ApJ...825...11C} on orbits with relative
semimajor axes $a_{\rm{rel}} \gtrsim 400$\,AU can be resolved using {\it
HST/WFC3}'s UVIS channel ($0.04$''/pixel, i.e., separations $\gtrsim 0.1$'' are resolved). Thus, the range of companion semimajor axes contributing this bias is $400 \lesssim a_{\rm{rel}} \lesssim 4000$\,AU. The threshold for resolving the widest LMC companions would be $a_{\rm{rel}} \gtrsim 10000$\,AU.

We estimate the bias due to wide binaries using published multiplicity statistics of intermediate-mass stars in conjunction with predictions from state-of-the-art stellar evolution models. We expect this to be small since a) wide binaries are rare and b) Cepheids outshine typical companions by several magnitudes \citep[e.g.][]{2016ApJS..226...18A}.
We rely on a  general compilation of intermediate-mass\footnote{The mass range of Cepheid progenitors with Solar metallicity is approximately $5-9\,\mathcal{M}_\odot$ \citep{2014A&A...564A.100A,2016A&A...591A...8A}} star multiplicity statistics \citep{2017ApJS..230...15M} rather than on Cepheid-specific multiplicity information, since the range of possible orbital semimajor axes\mdash in particular the very wide orbits\mdash has been more completely explored for B-stars than for the generally distant and evolved Cepheids. 
We further base the estimate of the typical wide companion flux on stellar model predictions, since empirical estimates of this kind are not presently available.

We first discuss the occurrence rate of wide companions, $f_{\rm{wb}}$, in \S\ref{sec:binaryfraction}, and then estimate the average flux contribution of a typical companion star, $\hat{M}_{P_p}$, in \S\ref{sec:typicalcompanionbias} to assess the possible impact on $H_0$.

\subsection{Wide binary fraction}
\label{sec:binaryfraction}

We adopt $f_{\rm{wb}} = 0.15$ as the fraction of MW Cepheids that have wide ($a_{\rm{rel}} > 400$\,au), spatially resolved, companions. This value is based on a recent comprehensive compilation of information on binary statistics that employed results from spectroscopy, eclipses, long-baseline interferometry, sparse aperture masking, adaptive optics, lucky imaging, and common proper motion \citep{2017ApJS..230...15M}. The de-biased multiplicity fraction of intermediate-mass stars with mass ratios $q = \mathcal{M}_2/\mathcal{M}_1 > 0.1$ in the orbital period range $\log{(P_o\ \rm{[d]})} \in [6.5,7.5]$ is $f_{\log{P_o}=[6.5,7.5];q>0.1} = 0.11\pm0.03$. We adopt a slightly higher $f_{\rm{wb}} = 0.15$ to account for a larger orbital period range of interest, i.e., $\log{(P_o\ \rm{[d]})} \in [7.0,8.5]$, for typical Cepheid companions with $a_{\rm{rel}} \in [400,4000]$\,AU and $\mathcal{M}_1 \in [5, 9]\,M_\odot$. 
Note that the adopted binary fraction is an upper limit in the
sense that $f_{\log{P_o}=[6.5,7.5];q>0.1}$ also contains mass ratios $q = \mathcal{M}_{\rm{Companion}} / \mathcal{M}_{\rm{Cep}} \in [0.1,
0.3]$ rather than the presently adopted $q > 0.3$. This limitation is related to
a) the limited mass range in the stellar isochrones and b) differences in the
mass function for lower $q$ \citep{2017ApJS..230...15M}.

\subsection{Typical companion bias based on stellar models}
\label{sec:typicalcompanionbias}
\label{sec:assum}

We estimate the light contribution by companions on wide orbits using stellar isochrones and random companion mass ratios, since the mass ratio distribution of very wide companions $(a_{\rm{rel}} \approx 200 - 5\,000 \rm{AU})$ is nearly consistent with random pairing following the initial mass function across the mass ratios of interest \citep{2017ApJS..230...15M}. 
We adopt a power law distributions, $p_q \propto q^{\gamma}$,
where $\gamma = -2$ for $q > 0.3$   for stars in the
mass range $5 - 9\,M_\odot$. We draw 100,000 random mass ratios using a
power law normalized such that there is a $100\%$ chance of having a companion
within $q = [0.3,1.0]$, i.e., $\int_{0.3}^{1.0} const \cdot q'^{-2} dq' = 1$.

\begin{figure}
\centering
\includegraphics{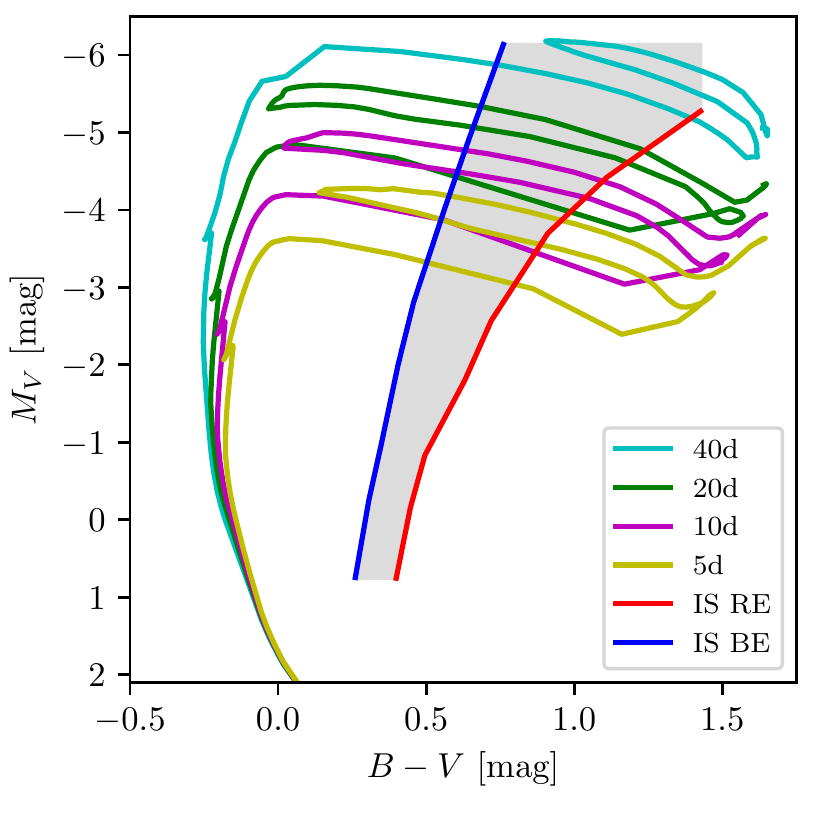}
\caption{Solar metallicity Geneva isochrones with average initial rotation rate ($\Omega/\Omega_c = 0.5$) corresponding to different pulsation periods near the blue instability strip edge during the second crossing. The red (IS RE) and blue (IS BE) edges of the instability strip shown are based on the same models \citep{2016A&A...591A...8A}.}
\label{fig:GVAisochrones}
\end{figure}

\begin{figure}
\centering
\includegraphics{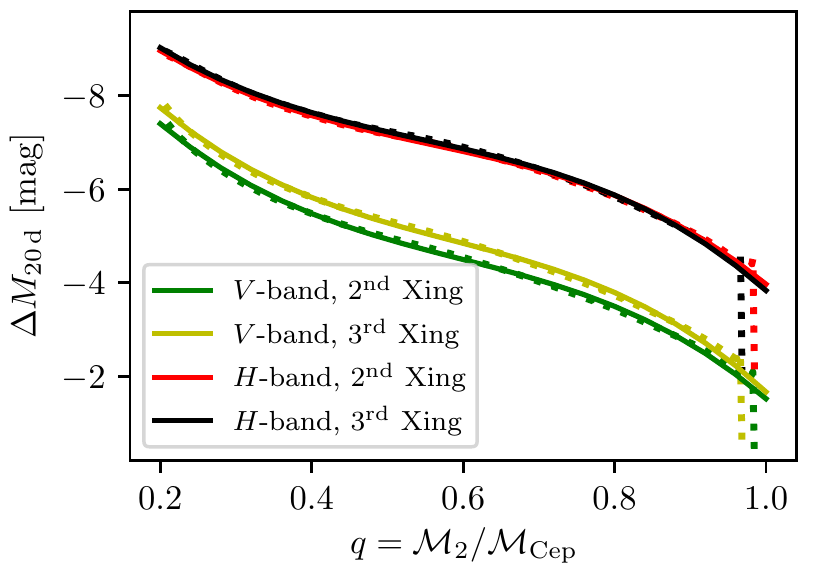}
\caption{Contrast between a 20\,d Cepheid and its Main Sequence companion as a function of mass ratio $q$. The corresponding isochrone is shown in Fig.\,\ref{fig:GVAisochrones}. The label `Xing' in the legend indicates second or third IS crossings.}
\label{fig:ContrastMassRatio}
\end{figure}

We estimate the typical flux contribution due to a wide binary companion as
follows. Predictions of stellar properties are provided by Geneva isochrones
\citep{2012A&A...537A.146E,2013A&A...553A..24G,2014A&A...564A.100A} of Solar
metallicity ($Z = 0.014$) computed using a typical initial (ZAMS) angular rotation rate
$\omega = \Omega / \Omega_{\rm{crit}} = 0.5$. 

For simplicity we assume that Cepheids are located near the hot edge of the
instability strip (IS) determined for these models \citep{2016A&A...591A...8A}.
Although real Cepheids are distributed across the IS, this simplifying
assumption eliminates the uncertainty
related to the location of the cool IS edge and provides a more conservative
upper limit on the companion bias, which tends to decrease for redder stars, in particular when dealing with reddening-free Wesenheit magnitudes. The blue edge of the IS was approximated by a linear fit:
$\log{(T_{\rm{eff,BE}}\,[\rm{K}])} = 3.9300 - 0.0447 \log{L/L_\odot}$.

 \begin{figure*}
  \centering
  \includegraphics{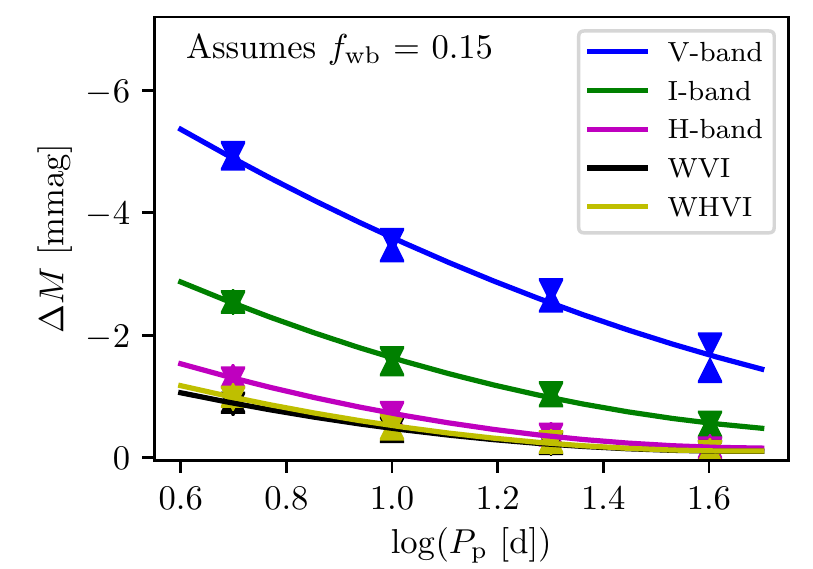}
  \includegraphics{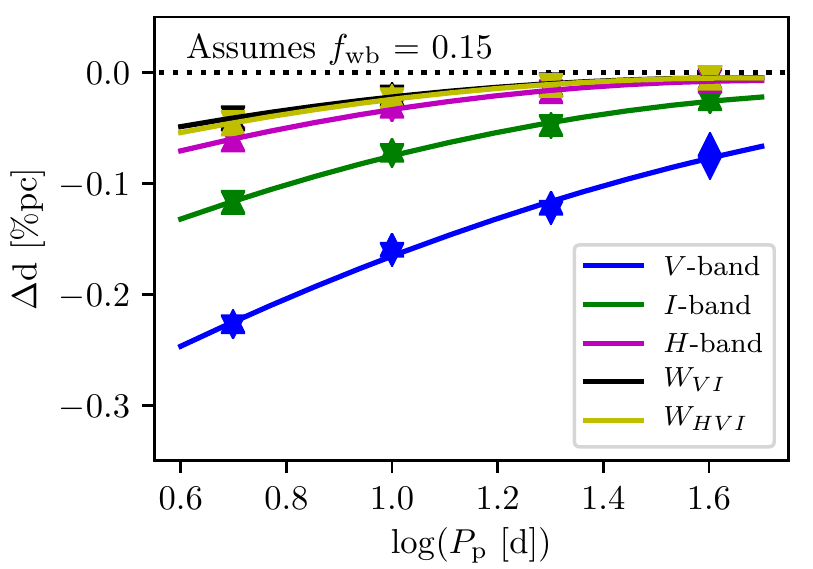}
  \caption{Estimation of bias due to wide binaries in apparent magnitude (left) and percent of distance (right). Negative $\Delta M$ indicates an apparent brightening due to additional companion flux, negative distance indicates that extragalactic distances have been underestimated by the stated amount. Both panels assume that $f_{\rm{wb}} = 15\%$ of Cepheids have companions that are resolved only on Galactic, but not on extragalactic scales.}
    \label{fig:magnitudebias}
\end{figure*}

The isochrones used for this estimation have been computed using a freely
accessible online interpolation
tool\footnote{\url{https://obswww.unige.ch/Recherche/evoldb/index/}}.
The adopted isochrone ages were computed using period-age relations determined
for the same models \citep{2016A&A...591A...8A} and correspond to a range of
pulsation periods of interest ($P_{\rm{p}} = [40, 20, 10, 5]\,$d).
Age differences between second and third IS crossings are accounted for, i.e., a
40\,d Cepheid on the second crossing is slightly younger than a 40\,d Cepheid on
the third crossing. Thus, the adopted isochrone ages are $\log{(t\,[\rm{yr}])}=$
7.36, 7.57, 7.78, and 7.99 for Cepheids on second crossings, and
$\log{(t\,[\rm{yr}])}=$ 7.41, 7.62, 7.83, 8.04 for third crossing Cepheids. The
isochrone files list VEGA magnitudes for UBVRI (Johnson-Cousins
photometric system) and JHK (Bessel) filters \citep{2016ApJS..226...18A}.
Detailed comparisons with Cepheid properties have shown excellent agreement for a host of different observables \citep{2016A&A...591A...8A}.

All companion stars considered here are assumed to be main sequence stars. This
implies a practical upper limit on the mass fraction $q = \mathcal{M}_2/\mathcal{M}_1 \lesssim
0.97$. Although five eclipsing LMC binary systems composed of Cepheid and red
giant stars have been identified by the {\it Araucaria} project
\citep[e.g.][]{2014ApJ...786...80G}, the short lifetime of the Cepheid
evolutionary stage \citep[$10^4 - 10^6$yr depending on mass ($9 - 5\,\mathcal{M}_\odot$), cf. Tab.\,4 in][]{2014A&A...564A.100A} in practice limits observed pairs of evolved binary components to the shortest
period Cepheids. To wit, the longest period Cepheid in such as system has
$P_{\rm{p}} = 3.8$\,d \citep{2010Natur.468..542P}. However, such short
periods are not usually observable in distant SN host galaxies and are  
not relevant to the present discussion. Similarly, we do not consider white dwarf or other stellar remnant companions, although it is known that the companions of approximately $30 \pm 10\%$ of single-lined spectroscopic binaries among O and B-type stars are stellar remnants, the majority of which are white dwarfs \citep{1978ApJ...222..556W,1980ApJ...242.1063G,2017ApJS..230...15M}. Neglecting the occurrence of white dwarf companions overestimates the binary bias, since white dwarf companions do not measurably affect Cepheid magnitudes, and more than compensates for neglecting the rare cases of binary systems containing a Cepheid and another evolved star. However, we here do not reduce $f_{\rm{wb}}$ to account for white dwarf companions due to a lack of detail in binary statistics. For instance, the fraction of white dwarf companions in very wide (isolated) binaries may be significantly different from single-lined spectroscopic binaries (SB1), which are more likely to undergo binary interactions. Any such correction will linearly affect the total estimated bias and can be straightforwardly applied by modifying $f_{\rm{wb}}$ according to the fraction of companions assumed to be stellar remnants.

For a fixed mass ratio, the luminosity contrast $\Delta M$ between a Cepheid and
its main sequence companion depends exclusively on the Cepheid's pulsation period $P_{\rm{p}}$ and IS
crossing, since initial rotation rate $\omega$ and metallicity $Z$ are assumed
to be identical, and the Cepheid's temperature is fixed to coincide with the 
blue IS edge.

For each isochrone, we compute magnitude differences between the Cepheid and a
range of companion masses with $0.3 < q < 0.95$ based on a cubic polynomial fit
in the $q$ vs. $\Delta M$ plane. This fit yields $\Delta M(\log{t};q) = \Delta
M(P_p;\rm{Xing};q) = M_{P_p;\rm{Xing}} - M_{\rm{Comp};q}$, where the first identity reaffirms that each isochrone age has been selected to match a specific combination of pulsation period and
crossing number. We compute the typical companion photometric bias as:
\begin{equation}
  \hat{M}_{P_p;\rm{Xing}} = \langle -2.5 \log{(1 + 10^{0.4 \cdot \Delta
  M_{P_p;\rm{Xing};q}})} \rangle \ ,
\end{equation}
where $\hat{M}_{P_p;\rm{Xing}}$ denotes the mean over the distribution computed using $100\,000$ values of $\Delta M(P_p;\rm{Xing};q)$ evaluated for random $q$. 
Since $q$ follows a power law \citep{2017ApJS..230...15M}, the distribution of
$\Delta M$ obtained from this simulation also follows a power law. Consequently,
the difference between the mean and median $\Delta M$ is approximately a factor of
two in all filters, with the mean giving a larger bias. Uncertainties in
the slope of the power law, etc., have been neglected for the time being.     

The typical photometric bias due to such wide companions translates to a fractional distance
bias as:
\begin{equation}
 d_{\rm{biased}}/d_{\rm{true}} = 10^{-0.2\cdot \hat{M}_{P_p;\rm{Xing}}}\ .
\end{equation}

\begin{table*}
  \centering
  \begin{tabular}{rrrrrrrrrrr}
    \hline
    \hline
    $P_p$ & \multicolumn{2}{c}{$V$-band} & \multicolumn{2}{c}{$I$-band} & \multicolumn{2}{c}{$H$-band} & \multicolumn{2}{c}{$W_{\rm{VI}}$} & \multicolumn{2}{c}{$W_{\rm{H,VI}}$} \\
    ~(d) & \multicolumn{2}{c}{(mmag)} & \multicolumn{2}{c}{(mmag)} & \multicolumn{2}{c}{(mmag)} & \multicolumn{2}{c}{(mmag)} & \multicolumn{2}{c}{(mmag)} \\
    & $2^{\rm{nd}}$ & $3^{\rm{rd}}$ & $2^{\rm{nd}}$ & $3^{\rm{rd}}$ & $2^{\rm{nd}}$ & $3^{\rm{rd}}$ & $2^{\rm{nd}}$ & $3^{\rm{rd}}$ & $2^{\rm{nd}}$ & $3^{\rm{rd}}$ \\ 
    \hline
    \multicolumn{11}{c}{For a Cepheid with a typical wide companion:}\\
    5 & -33.0 & -32.4 & -16.8 & -16.9 & -8.6 & -8.8 & -5.8 & -6.1 & -6.5 & -6.7 \\
    10 & -23.5 & -22.4 & -10.7 & -10.1 & -4.8 & -4.4 & -3.2 & -2.9 & -3.5 & -3.2 \\
    20 & -18.1 & -17.0 & -7.0 & -6.8 & -2.5 & -2.5 & -1.6 & -1.6 & -1.7 & -1.7 \\
    40 & -12.2 & -9.4 & -3.8 & -3.4 & -1.0 & -1.2 & -0.6 & -0.7 & -0.6 & -0.8 \\

    \hline
    \multicolumn{11}{c}{Taking into account the wide binary fraction, $f_{\rm{wb}} = 0.15$:}\\
    5 & -4.9 & -4.9 & -2.5 & -2.5 & -1.3 & -1.3 & -0.9 & -0.9 & -1.0 & -1.0 \\
    10 & -3.5 & -3.4 & -1.6 & -1.5 & -0.7 & -0.7 & -0.5 & -0.4 & -0.5 & -0.5 \\
    20 & -2.7 & -2.5 & -1.0 & -1.0 & -0.4 & -0.4 & -0.2 & -0.2 & -0.2 & -0.3 \\
    40 & -1.8 & -1.4 & -0.6 & -0.5 & -0.1 & -0.2 & -0.1 & -0.1 & -0.1 & -0.1 \\
    \hline
    \hline
  \end{tabular}
  \caption{Mean photometric bias due to wide binaries. For each band and filter
  combination, values for second and third crossings are shown separately.
  $W_{\rm{VI}} = I - 1.55\cdot(V-I)$ and $W_{\rm{H,VI}} = H - 0.4\cdot(V-I)$ are
  the reddening-free ``Wesenheit'' magnitudes 
  \citep{1982ApJ...253..575M,2008AcA....58..163S,2011ApJ...730..119R} assuming $R_V = 3.1$ \citep{1989ApJ...345..245C}.      
    \label{tab:magnitudebias}}
\end{table*}

Table\,\ref{tab:magnitudebias} summarizes our results obtained for different
Cepheid pulsation periods and IS crossings, in $V$, $I$, and $H-$band as well as Wesenheit formulations. To estimate the influence on $H_0$, these individual distance bias estimates must be multiplied by the fraction of stars concerned:
\begin{equation}
\hat{M}_{P_{\rm{p;wb}}} = f_{\rm{wb}} \cdot \hat{M}_{P_{\rm{p}}}\ .
\label{eq:compbiasinclfrac}
\end{equation}

Table\,\ref{tab:magnitudebias} and Figure\,\ref{fig:magnitudebias} present the (very small) effect in terms of magnitude and percent distance taking into account the wide binary fraction. 
The largest effect is found in $V-$band for the shortest-period Cepheids (here: $5$\,d).
In this worst case scenario, the bias amounts to approximately $-5\,$mmag, or a
distance error of $-0.23\%$. Focusing on Cepheids with $P_{\rm{p}} > 10$\,d,
which are most relevant for extragalactic applications, and on near-IR
photometry or Wesenheit magnitudes reduces this
effect to the sub-mmag level and to fractional distance errors smaller than
$0.02\%$. Since the MW is the only {\it SH0ES} anchor galaxy in which wide binaries
may be resolved and is 1 of 3 equivalently weighted anchors, the impact of this bias enters the $H_0$ estimation weighted by approximately $1/3$. For a typical pulsation
period of $20$\,d and using NIR-based Wesenheit formulation, the distance effect
due to such wide companions is a mere $4 \cdot 10^{-5} = 0.004\%$, which is clearly
negligible, even en route to $1\%$ $H_0$ accuracy.

\section{Cepheids in Open Clusters}
\label{sec:ClusterCepheids}

To assess the impact of a cluster-related bias, we proceed as follows. We
first estimate the occurrence rate of Cepheids in clusters by reviewing the literature for cluster Cepheids in the MW, LMC, SMC, and M31, cf.
\S\ref{sec:CCfrequency}. We then measure the average bias arising
from cluster populations using observations of nine M31 cluster Cepheids, cf. \S\ref{sec:M31ClusterCepheidPhotometry}. We finally estimate the influence of
cluster-related stellar association bias on $H_0$ in \S\ref{sec:CC-H0}.

\subsection{Occurrence rate of Cepheids in clusters}
\label{sec:CCfrequency}

Several previous studies aimed at investigating the effects of blending on
Cepheid photometry and the distance scale have (incorrectly) asserted that Cepheids very frequently
occur in open clusters based on the notion that young stars tend to reside in clusters. However, most embedded and gravitationally unbound star clusters or associations disperse within a few to a few tens of Myr \citep[$10\%$ of embedded clusters with mass exceeding $150\,\mathcal{M}_\odot$ survive for 10\,Myr or more]{2003ARA&A..41...57L,2006MNRAS.373..752G,2012MNRAS.425..450M}. For comparison, Cepheids are several tens to a few hundred Myr old \citep{2016A&A...591A...8A}. Here we determine the occurrence rate of Cepheids in clusters, which is crucial for correctly evaluating the impact of cluster-related blending on the distance scale.  

We estimate the clustered fraction of fundamental mode Cepheids in the MW, $f_{\rm{CC,MW}}$, using Cepheids that are bona fide members of Galactic open clusters located within $2$\,kpc of the Sun. The sample of bona fide cluster Cepheids is taken from \citet[henceforth: A+13]{2013MNRAS.434.2238A} and contains a total of 11 objects\footnote{Ordered by distance, this list of bona fide cluster Cepheids consists of (not counting inconclusive or unlikely candidates from A+13): Y~Sgr, U~Sgr, V~Cen, SU~Cyg, BB~Sgr, S~Mus, S~Nor, X~Cyg, CV~Mon, DL~Cas, SX~Car, UW~Car, and RU~Sct.}. The {\it Gaia} mission will soon enable a revision of this estimate, which is purely for comparison with other estimates presented in the following.

We construct the reference sample of all fundamental mode Cepheids within $d < 2$\,kpc using the compilation of data on MW Cepheids used by A+13. This list comprises 130 Cepheids and is included in appendix \ref{app:MWCeps}. The reference sample notably includes several new candidate Cepheids from the ASAS survey \citep{2002AcA....52..397P,2003AcA....53..341P,2004AcA....54..153P,2005AcA....55...97P} whose classification was confirmed using spectroscopic observations \citep[Sec.\,2.4]{2013PhDT.......363A}. We adopt average distances from the A+13 compilation and the McMaster Cepheid database\footnote{\url{http://www.astro.utoronto.ca/DDO/research/cepheids/cepheids.html}} \citep{1995IBVS.4148....1F} and perform a hard cut off (i.e., using the computed distances at face value) at $2$\,kpc. We limit this estimate to heliocentric distances $d<2$\,kpc because of the apparent decrease in MW cluster detection efficiency (see Figure 20 in A+13). Pulsation modes are adopted from \citet{2009A&A...504..959K}, except where $P_{\rm{p}} > 7.57$\,d, in which case we assume FU mode pulsation \citep[i.e., we adopt V440~Per as the overtone Cepheid with the longest period following][]{2009MNRAS.396.2194B}.

We thus find $f_{\rm{CC,MW}} = 11/130 = 8.5\%$ for fundamental mode Cepheids within $2$\,kpc of the Sun. Further considerations regarding this fraction are presented in \S\ref{sec:disc:clusters}.

There are two well studied LMC clusters that contain up to 24 classical Cepheids each:
NGC\,1866 and NGC\,2031
\citep{1993AJ....105.1813W,2007A&A...462..599T,2016MNRAS.457.3084M}. These impressive cases contain up to 8 times the number of Cepheids found in the record-holding MW cluster NGC\,7790 with its 3 member Cepheids \citep[A+13]{1958ApJ...128..150S} and have thus received much attention. However, it should be noted that there is no known equivalent for such clusters in the MW, the SMC, or M31. Moreover, these well-known special cases are also not
representative of the typical Cepheid host clusters in the Magellanic
Clouds. The most comprehensive survey of cluster Cepheids in the LMC and SMC to date
\citep{1999AcA....49..543P} indicates that the vast majority of Cepheid-hosting
clusters in both MCs contain only one or two Cepheids, which is similar to the
MW. 

Using the available data for LMC and SMC cluster Cepheids based on phase two of the Optical Gravitational Lensing Experiment \citep[{\it OGLE-II}]{1997AcA....47..319U,1999AcA....49..223U,1999AcA....49..437U,1999AcA....49..543P}, we determine the following clustered Cepheid fractions, omitting clusters containing more than 3 Cepheids: $f_{\rm{CC,LMC}} = 81 / 740 = 11\%$ and $f_{\rm{CC,SMC}} = 76/1272 = 6.0\%$. These numbers illustrate that the LMC and SMC cluster Cepheid fraction are not so different from the MW. Specifically for long-period Cepheids ($P_{\rm{p}} > 10$\,d), the clustered fraction is $f_{\rm{CC,LMC,longP}} = 4/55 = 7.2\%$ in the LMC and $f_{\rm{CC,SMC,longP}} =  6/74 = 8.1\%$. While these numbers would benefit from an update based on the now completed survey of Cepheids in the Magellanic system \citep[{\it OGLE-IV}]{2017AcA....67..103S}, it is obvious that the occurrence of cluster Cepheids is also comparatively uncommon in the Magellanic clouds.

\begin{figure}
\centering
\includegraphics{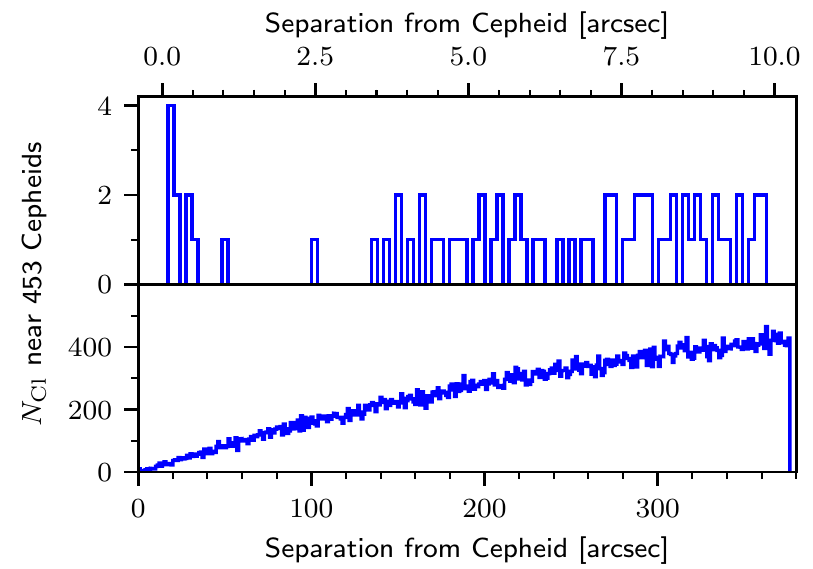}
\caption{Number of M31 star clusters cross-matched within a given separation near M31 Cepheids (PS1 sample). The 11 known cluster Cepheids in M31 \citep{2015ApJ...813...31S} have separations $\lesssim 1.0"$ from nearby cluster centers. A total of $535$ PS1 Cepheids are found within the PHAT footprint, of which $453$ were  classified as neither type-II nor first overtone Cepheids. The upper panel shows a histogram for the number of clusters within separations smaller than $10$\,arcsec. The bottom panel shows the cumulative histogram of AP clusters situated at separations closer than $6.3'$ (100 {\it WFC/IR} pixels at $22.9$\,Mpc) from these Cepheids.}
\label{fig:ClnearCepM31}
\end{figure}

The Andromeda galaxy (M31) provides a useful analog for both the MW and the {\it SH0ES} project's SN-host galaxies \citep{2016ApJ...830...10H}, all of which are relatively high-mass spiral galaxies with similar (high) metallicity. To wit, $\log{(\mathcal{M}_{\rm{stars}}/\Msol)} = 10.80$ (MW) and $11.1$ (M31) \citep{2011MNRAS.414.2446M,2012A&A...546A...4T} compare to the mean stellar mass of the {\it SH0ES} galaxies ($\log(\mathcal{M}_{stars}/\Msol) = 9.8$). Moreover, the average oxygen abundance, which is commonly used as a proxy for metallicity in distance scale matters \citep{1998ApJ...498..181K}, is $12+\log\rm{(O/H)}=8.9$ for each of the MW, M31, and NGC\,4258, whereas it is $8.91$ for the {\it SH0ES} galaxies (mean of WFC3 IR Cepheids in tab.\,4 of R+16). For comparison, the LMC has even lower mass \citep[$\log(\mathcal{M}_{stars}/\Msol) = 9.4$]{1998ApJ...503..674K} and metallicity ($12+\log\rm{(O/H)}=8.65$), which results in distinct differences in its Cepheid populations, such as shorter minimum periods \citep[e.g.][]{2016A&A...591A...8A,2017EPJWC.15206002A}. M31 furthermore provides an external view on its stellar populations and has been imaged extensively using {\it HST}, which renders M31 a unique laboratory for estimating stellar association bias as it applies to the cosmic distance scale.

The population of Cepheids in M31 is well known thanks to the Pan-STARRS survey \citep[PS1]{2013AJ....145..106K}. High-quality deep {\it HST} imaging in multiple passbands of approximately one third of M31's disk is available via the Panchromatic Hubble Andromeda
Treasury project \citep[PHAT]{2012ApJS..200...18D}. M31 star clusters have been identified by the Andromeda Project \citep[AP]{2015ApJ...802..127J}, and 11 Cepheid-hosting clusters have been reported \citep{2015ApJ...813...31S}, cf. Tab.\,\ref{tab:CCsample}.
Cross-matching all PS1 Cepheids with the PHAT source catalog \citep{2014ApJS..215....9W}, we identify $535$ PHAT sources within angular separations of typically less than $0.2"$ (maximum separation $1.2"$) from the PS1 input positions. Of these, $365$ are classified as fundamental mode (FM) Cepheids, $88$ are unclassified (UN), and $82$ are either classified as first overtone or as type-II Cepheids.
Given the 11 M31 cluster Cepheids, we find $f_{\rm{CC,M31}} = 11/453 = 2.4\%$ when including unclassified Cepheids and $9/365 = 2.5\%$ when using only PS1 objects classified as fundamental mode Cepheids. 

\begin{figure}
\centering
\includegraphics{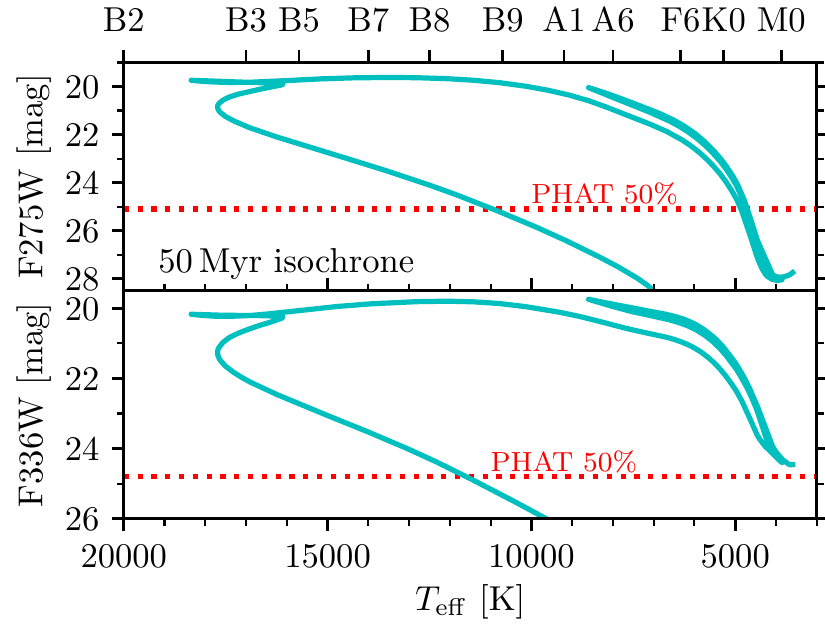}
\caption{Predicted UV magnitudes at M31's distance ($\mu_{\rm{M31}}=24.36$\,mag as adopted in R+16) against temperature and spectral type. A $50$\,Myr Solar metallicity PARSEC isochrone was computed using the {\it HST/WFC3} UVIS passbands F275W and F336W. Predicted $T_{\rm{eff}}$ values are translated to spectral types using the empirical calibration
by \citet{2013ApJS..208....9P}. The $50\%$ completeness level \citep[from Fig.\,12 in][]{2012ApJS..200...18D} corresponds to a late B spectral type and a stellar mass of approximately $2.5\,M_\odot$. UV observations of Cepheids are particularly well suited to detect host cluster populations. 
\label{fig:UVclusterdetectability}}
\end{figure}

To ensure that these cluster counts were complete in the vicinity of the Cepheids, we decided to visually inspect UV (F275W and F336W filters) postage stamps similar to Fig.\,\ref{fig:CCpostagestamps} for all 453 FM or UN PS1 Cepheids for the presence of any additional clusters that were not identified by the AP. Figure\,\ref{fig:UVclusterdetectability} uses the PHAT project's completeness estimations \citep[cf. Fig.\,12 in][]{2012ApJS..200...18D} and a comprehensive listing\footnote{\url{http://www.pas.rochester.edu/~emamajek/EEM_dwarf_UBVIJHK_colors_Teff.txt}} of spectral types and temperatures \citep{2013ApJS..208....9P} to illustrate that F275W and F336W  observations are highly suitable for detecting the vast majority of M31 clusters containing B-stars. 
The inspected UV postage stamps are provided in the online appendix\,\ref{app:postagestamps} for the reader's convenience and reproducibility. 
The inspection of the UV postage stamps thus confirms that the overall clustered Cepheid fraction within the PHAT footprint is a factor of 3 to 4 lower than the MW and LMC/SMC fractions established above. Further considerations regarding completeness of $f_{\rm{CC,M31}}$ are presented in \S\ref{sec:disc:averagebias}.

Each of the aforementioned estimations of $f_{\rm{CC}}$ are subject to their own specific systematics and uncertainties. However, variations of $f_{\rm{CC}}$ among galaxies are to be expected because of 
differences in cluster dispersal timescales that depend on various factors, including galactic potentials and the presence of molecular clouds. Notably for the MW, establishing a galaxy-wide average including long-period Cepheids is complicated by the detection of clusters against the foreground in the Galactic disk and the rare occurrence of long-period Cepheids near the Sun. In the case of the Magellanic clouds, a collision may have happened roughly $100-300$\,Myr ago \citep{2012MNRAS.421.2109B}, i.e., during the time when many of the presently observable Cepheids were born. However, it is reasonable to expect a clustered Cepheid fraction in the lower percent range, given the $\sim 10$\,Myr timescale for cluster dissociation and typical  ages of $50-75$\,Myr for $20-10$\,d Cepheids.

In the following, we adopt $f_{\rm{CC,M31}} = 2.5\%$  for estimating the impact of stellar association bias on $H_0$. This choice is based on three main arguments. First, $f_{\rm{CC,M31}}$ is consistent with (by construction) the average cluster bias measured on M31 data in \S\ref{sec:M31ClusterCepheidPhotometry}. Second, M31 is more alike SN host galaxies than the Magellanic Clouds with their potentially peculiar star formation histories and low metallicities, and provides an external view similar to distant galaxies that is not available for the MW. Third, the impact of cluster contamination on $H_0$ actually depends on the \emph{effective} value of $f_{\rm{CC}}$ as it applies to SN host galaxies across the cosmic distance ladder, and not necessarily on the \emph{true} value of $f_{\rm{CC}}$. \S\ref{sec:disc:clusters} discusses this point in detail and shows that the M31 cluster population is highly suitable for estimating an upper limit on stellar association bias due to clusters. \S\ref{sec:disc:fcc} considers further elements affecting the \emph{true} fractions of Cepheids occurring in clusters.

\begin{figure}
\centering
\includegraphics{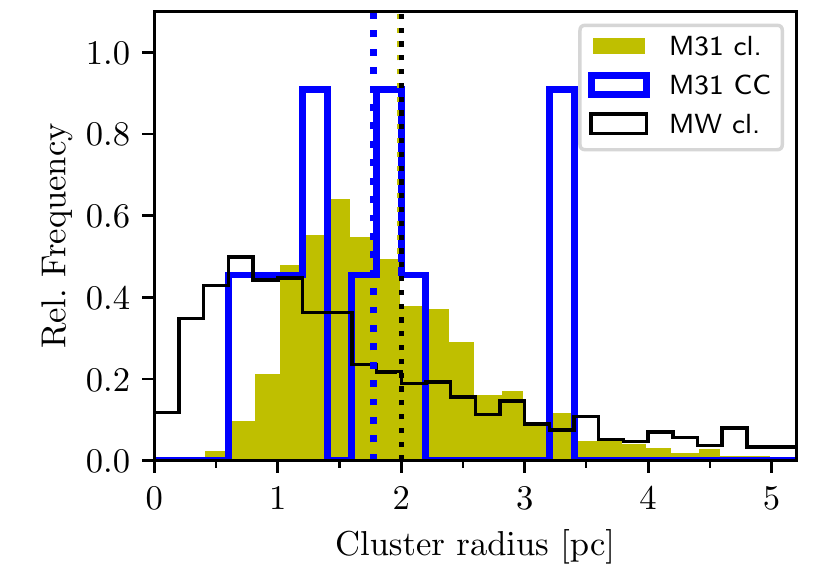}
\caption{Relative frequency of cluster radii in M31
\citep[yellow area]{2015ApJ...802..127J} and the Milky Way
\citep[black line, restricted to ones within $2$\,kpc
of the Sun]{2002A&A...389..871D}, and the relative frequency of cluster radii
among Cepheid host clusters in M31
\citep[thick blue line]{2015ApJ...813...31S}. The mean cluster radii in M31 and
the MW are virtually identical, $\langle r_{\rm{cl}}\rangle = 2.0$\,pc. The
mean radius of M31 Cepheid host clusters is $1.8$\,pc.}
\label{fig:clusterrads}
\end{figure}

\subsection{Measuring Cluster Cepheid Bias in M31} 
\label{sec:M31ClusterCepheidPhotometry}

\begin{figure}
\centering
\includegraphics{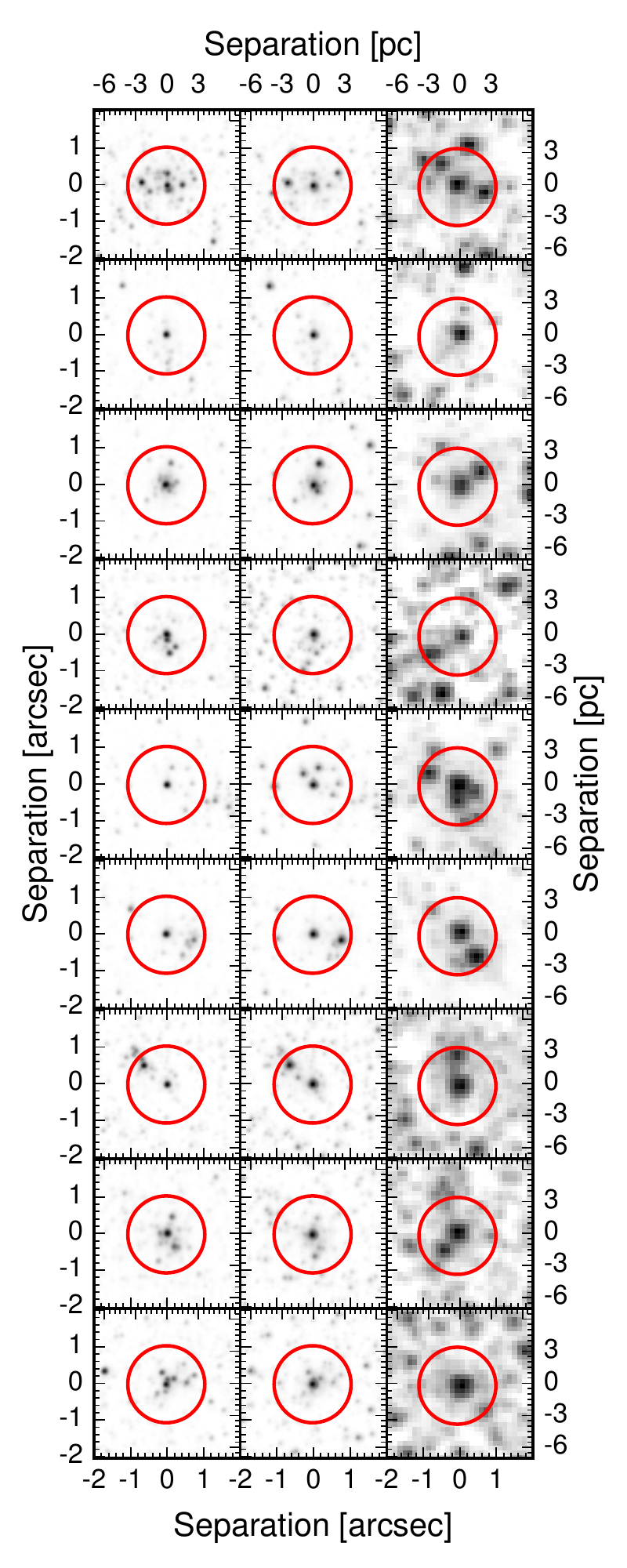}
\vspace{-0.5cm}
\caption{Postage stamps of the 9 analyzed cluster Cepheids in
M31. For every star (sorted by increasing $P_{\rm{p}}$ and M31CC-ID, cf. Tab.\,\ref{tab:CCsample}), we show
cutouts from the F475W, F814W, and F160W images, with wavelength increasing from
left to right. The red circle indicates the area over which the cluster
contributes light and has radius $r_{\rm{apt}} = 1.05''$. Note that F475W and
F814W have identical orientation and a finer plate scale ($0.05$''/pix ACS
compared to WFC3/IR channel $0.13$''/pix, which is also rotated.)} 
\label{fig:CCpostagestamps}
\end{figure}

In contrast to the synthetic approach adopted for binaries, we estimate the stellar
association bias due to clusters empirically using observations of M31 obtained
via the PHAT project \citep{2012ApJS..200...18D}. M31 and PHAT data 
provide a suitable empirical base for this estimation, since clusters in M31 are
sufficiently spatially resolved, optical and near-IR images are available in the
{\it HST} filter system, and M31 is an appropriate analog for comparison
with SN-host galaxies. Moreover, basing this estimate on  
observations implies that certain typical observational systematics such as overlapping stellar populations, variable background, and other noise sources are naturally included in the estimation.

\begin{table*}
\centering
\begin{tabular}{lrrrrr}
\hline
\hline
M31CC-ID & Cepheid PS1 ID & AP ID & $P_{\rm{p}}$ & $a_{\rm{Cep}}$ &
$p_{\rm{Cl}}$ \\ 
 & & & (d) & (Myr) &  \\
\hline 
M31CC-01 & PSO-J011.4286+41.9275 & 477  & 4.582 & 122 &  0.99  \\
M31CC-02 & PSO-J011.6227+41.9637 & 3928 & 6.213 & 101 & 0.87   \\
M31CC-03 & PSO-J011.6374+42.1393 & 2831 & 7.928 & 88 & 0.91    \\
M31CC-04 & PSO-J010.9769+41.6171 & 3050 & 8.829 & 82 &  0.85   \\
M31CC-05 & PSO-J011.6519+42.1286 & 2113 & 15.429 & 59 & 0.98   \\
M31CC-06 & PSO-J011.7119+42.0449 & 2587 & 17.180 & 56 & 0.83   \\
M31CC-07 & PSO-J011.0209+41.3162 & 2967 & 19.566 & 51 & 0.66   \\
M31CC-08 & PSO-J011.2797+41.6217 & 1082 & 26.499 & 43 & 0.91   \\
M31CC-09 & PSO-J011.3077+41.8500 & 1540 & 35.750 & 36 & 0.98   \\
\hline
\multicolumn{6}{l}{Known M31 cluster Cepheids that could not be studied, see
text}\\
near {\it HST/ACS} chip gap  & PSO-J011.4279+41.8642 & 1539 & 6.080 & 103 & 0.94 \\
no F160W observation  & PSO-J011.0696+41.8571 & 5216 & 14.353 & 62 & 0.70 \\
\hline
\hline
\end{tabular}
\caption{Table of M31 cluster Cepheids \citep{2015ApJ...813...31S}
included in this study. We present a short-hand identifier in addition to the
original Cepheid \citep[PS1]{2013AJ....145..106K} and cluster identifiers
\citep[Andromeda Cluster Project]{2015ApJ...802..127J}. $P_{\rm{p}}$
denotes the PS1 pulsation period. Ages are inferred using period-age
relations for solar-metallicity Cepheid models that include rotation
\citep{2016A&A...591A...8A}. Column $p_{\rm{Cl}}$ represents a 
probability of the cluster existing (cf. Andromeda Project); this is however not
a cluster membership probability for the Cepheid.} 
\label{tab:CCsample}
\end{table*}

To measure the blending bias due to clusters, we downloaded the original PHAT
observations from MAST\footnote{\url{http://mast.stsci.edu}} and proceeded as
follows. Note that we have measured photometry ourselves, and did not rely
on published magnitudes by \citet{2014ApJS..215....9W}.

Nine of the eleven known M31 cluster Cepheids have been
identified as fundamental-mode pulsators \citep{2013AJ....145..106K}, whereas
the pulsation mode was not identified for two others.
One Cepheid with unidentified pulsation mode, PSO-J011.6374+42.1393, is likely
pulsating in the fundamental mode (\Ppuls = $7.928$\,d), since the longest-known
period of an overtone pulsator in the MW is $7.57$\,d
\citep{2009MNRAS.396.2194B}.
The other Cepheid with unidentified pulsation mode, PSO-J011.4279+41.8642
(\Ppuls = $6.08$\,d) was located too close to the ACS chip gap to measure
photometry using extended apertures described below.
Unfortunately, PSO-J011.0696+41.8571 (\Ppuls = $14.353$\,d), was not observed in F160W.
Thus, we have included 9 Cepheids for estimating the typical blending
due to M31 clusters. This sample is listed in Tab.\,\ref{tab:CCsample} together
with identifiers adopted for the present work that are ordered by pulsation
period.

Correctly identifying the Cepheids in the HST images is crucial, since their
astrometry is based on the lower spatial resolution of the PanSTARRS
survey ($0.258$"/pixel). To this end, we assumed that the Cepheid should be the brightest object within $1"$ from the PS1 Cepheid's position in the F814W image. This approach is a reasonable assumption due to the evolved state and yellow color of the
Cepheid and has been shown to yield a narrow near-IR M31 PLR with low dispersion \citep{2012ApJ...745..156R}. Nearby red giants of similar luminosity may rival or outshine the
Cepheid in F160W, whereas hot stars may approach the Cepheid's brightness in
F475W. We inspected all F475W, F814W, and F160W postage stamp cut outs centered
on the Cepheid positions to ensure that the relative contrast of nearby stars
would behave as expected for a Cepheid. Figure\,\ref{fig:CCpostagestamps} shows
these postage stamps.

\subsubsection{Measuring photometric bias due to clusters}
\label{sec:CCPhotBias}

There are three separate contributions to the flux in an aperture centered on a cluster Cepheid: 1) the Cepheid flux, 2) the resolved cluster flux, and 3) the average unassociated background. To estimate the isolated resolved cluster flux, we first measure and remove the curve of growth for point sources matched to a small, r=1 pixel, aperture centered on the Cepheid. The background is then statistically measured as the mean of a larger-than-the-cluster annulus, and subtracted from the curve of growth. This leaves an estimate of the cluster flux.

We measure each image's curve of growth $m_{2\rm{px}} - m_{N\rm{px}}$ using a series of apertures of increasing radius, starting from $r=1$ pixel to $r=1.5"$ (5 pc) for $\sim
100$ stars. The $5$\,pc radius was inspired by typical cluster radii, cf.
Fig.\,\ref{fig:clusterrads} that is based on the catalogs by \citet{2002A&A...389..871D,2015ApJ...802..127J}, and
the feasibility of measuring photometry in extended apertures in the PHAT images. Figure\,\ref{fig:clusterrads} shows the distribution of physical sizes of MW and M31 clusters, with the M31 Cepheid-hosting clusters highlighted. The apparent difference in cluster radii is likely due to the inhomogeneous definition of cluster radii in the Galactic catalog \citep{2002A&A...389..871D}, whereas a consistent definition applies to M31 clusters. The mean cluster radius of both distribution is $\sim 2$pc, which translates to angular sizes of $0.53$", $0.054$", $0.018$", and $0.010$" at the distances corresponding to M31, NGC\,4258, the average  of the {\it SH0ES} SN-host galaxies in R+16 ($22.9$\,Mpc), and the farthest R+16 galaxy: UGC\,9391 ($38.5$\,Mpc).

We define the sky background as the {\it mean} flux in an annulus just outside
the largest aperture used to measure the curve of growth ($1.5"$ to $2.5"$). We employ the mean rather than the median in
order to be sensitive to background stars in addition to the sky level, which in
itself would be probed more robustly via the median.
We then estimate the mean flux level of the Cepheid plus cluster stars within
the aperture by subtracting the mean flux of sources and non-sources in the
outer annulus from the aperture region containing the Cepheid and cluster stars.
This somewhat noisy estimate of the mean Cepheid plus cluster flux level depends
on the brightness and number of stars that lie within the aperture or annulus. An
additional source of noise is the random phase magnitude estimate of the
Cepheid, for which we do not correct here. 

To isolate the extra flux contributed
by cluster stars (without the Cepheid), we subtract the curve of growth measured
within the series of apertures from the mean Cepheid plus cluster flux. This
provides us with the cluster flux contributions as a function of angular separation from the Cepheid in each of the photometric passbands considered.

\begin{figure}
\centering
\includegraphics{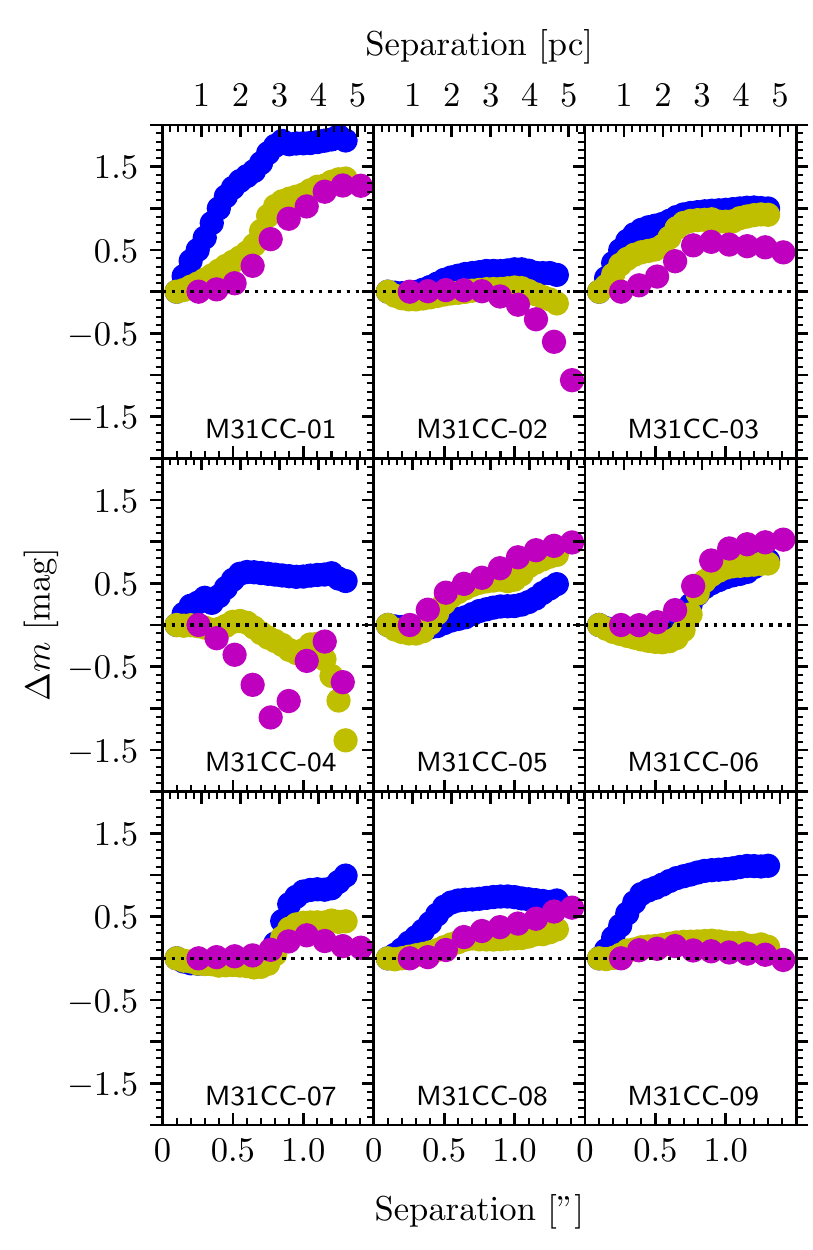}
\caption{Cumulative light contribution due to cluster stars for each M31 cluster Cepheid considered, cf.\,\S\ref{sec:M31ClusterCepheidPhotometry}. Colors represent different filters: F475W (blue), F814W (yellow), F160W (magenta). }
\label{fig:COGcluster}
\end{figure}

Figure\,\ref{fig:COGcluster} shows the resulting estimate of the cumulative
cluster flux contribution as a function of separation from the Cepheid for each
of the 9 cluster Cepheids. We have defined the blending bias for each
filter as
\begin{equation}
\centering
\Delta m = m_{\rm{Cl + Cep}} - m_{\rm{Cep}}\,,
\end{equation}
where $m$ denotes the magnitude in the filter of interest and $\Delta m$ is
negative when a light contribution beyond the background is present. Table
\ref{tab:M31deltamag} lists the inferred biases for for all 9 Cepheids. Note that the estimate of the cluster flux can be negative (positive $\Delta m$) due to its statistical nature, e.g., if the cluster flux is small and its location statistically sparser than the nearby (annulus) environment.

Figure\,\ref{fig:COGcluster} shows that $\Delta m_{\rm{F475W}}$ is negative at
most cluster radius apertures with a mean (median) of $-0.79 \pm 0.15\
(-0.72)$\,mag at an aperture corresponding to $r_{\rm{apt}} = 3.8$\,pc (cf.
below).
This agrees very well with the expectation that young clusters contain many hot
stars. In F814W the mean (median) $\Delta m_{\rm{F814W}} = -0.45 \pm 0.14\ (-0.43)$\,mag. For F160W, the mean (median) $\Delta m_{\rm{F160W}} = -0.39 \pm 0.16\ (-0.43)$\,mag, which is similar to the average bias in F814W. 
Again, the postage stamps (Fig.\,\ref{fig:CCpostagestamps}) illustrate the blue
nature of the (very centrally concentrated) cluster population dominated by hot
stars, whereas the field red giants (more widely distributed) become dominant
sources in F160W.

\begin{figure}
\centering
\includegraphics{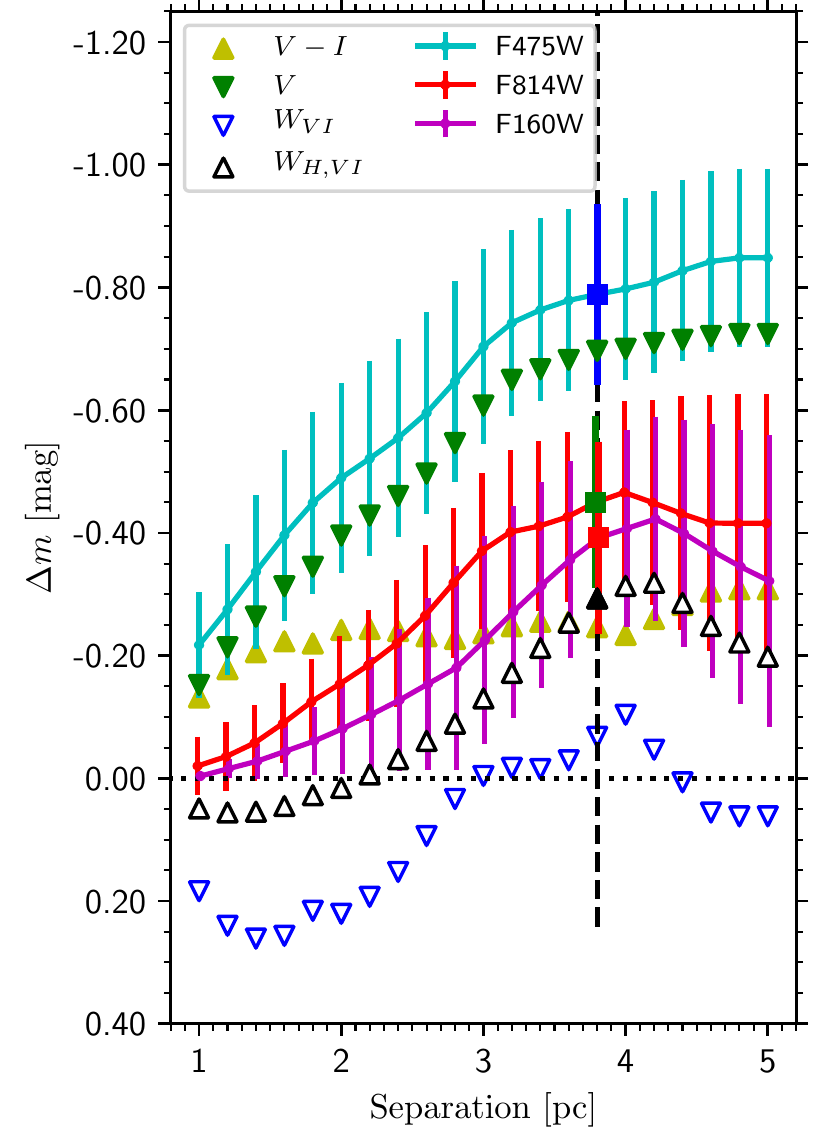}
\caption{Mean bias vs separation as obtained from the 9 individual cluster
Cepheids shown in Fig.\,\ref{fig:COGcluster}. $W$ denotes reddening-free Wesenheit magnitudes computed using optical photometry ($W_{VI}$) as well as combined near-IR and optical photometry ($W_{H,VI}$), cf.\,\S\ref{sec:M31ClusterCepheidPhotometry}. The dashed vertical line shows the separation $r_{\rm{apt}}=3.8$\,pc at which the cluster contribution levels off. Differently colored larger symbols are drawn here to illustrate that this
$\Delta m$ is used to evaluate stellar association bias due to clusters. $V-I$ color is estimated by color-color (F475W-F814W vs F555W-F814W) interpolation using a 50\,Myr solar-metallicity isochrone. The run of $V-I$ (yellow upward triangles) suggests evidence of mass segregation in the young Cepheid-hosting clusters. Intriguingly, optical Wesenheit magnitudes appear to be the least sensitive to the cluster light contribution. Combined near-IR and optical Wesenheit magnitudes are slightly less sensitive to cluster light contributions than near-IR data alone.} 
\label{fig:MeanBiasPerPc}
\end{figure}

Figure\,\ref{fig:MeanBiasPerPc} shows the mean cluster light contribution as a
function of separation obtained averaging the 9 curves of growth shown in Fig.\,\ref{fig:COGcluster}. The averaging here and in the following is done in magnitude space, i.e., we interpret the cluster contribution to scale with the luminosity of the Cepheid, cf. \S\ref{sec:disc:averagebias}. On average, noisy field star contamination cancels out and we recover a smoothly increasing cluster light contribution as a function of separation. From this average curve of growth, we find that the
cluster light contribution flattens off at a separation of about $r_{\rm{apt}} =
1.05" \equiv 3.8$\,pc. This distance corresponds to approximately twice the  average cluster radius of M31 open clusters, cf.
Fig.\,\ref{fig:clusterrads}. We thus adopt $r_{\rm{apt}} = 3.8\,$pc as the
typical area over which a cluster contributes light to a Cepheid's
photometry. Table\,\ref{tab:M31deltamag} lists the mean bias estimates at $r_{\rm{apt}}
= 3.8\,$pc.

We compute reddening-free Wesenheit magnitudes \citep{1982ApJ...253..575M}, using combinations of optical as well as optical and near-IR photometry. The F160W observations yield $H$-band magnitudes. We estimate $V-I$ color by interpolation using observed F475W-F814W colors and the color-color (F475W-F814W vs. F555W-F814W) space defined by a PARSEC isochrone\footnote{\url{http://stev.oapd.inaf.it/cmd}}
\citep{2017ApJ...835...77M} computed for an age of $50$\,Myr and
solar metallicity, since the PHAT project did not include observations in F555W ($V-$band).

Table\,\ref{tab:M31deltamag} lists the measured values of $\Delta m$\hbox{---}i.e., the change in apparent magnitude due to extra flux\hbox{---}for each
cluster Cepheid and passband or Wesenheit magnitude separately, as well as the (magnitude) mean of all 9 cluster Cepheids including standard error and median. 
Intriguingly, we find that the bias estimate for the optical Wesenheit magnitude $W_{VI}$ is the lowest among the filter combinations studied, since the integrated cluster contribution in $V-I$ multiplied by the reddening law partially compensates the $I$-band bias. Moreover, we find evidence of mass segregation in the young Cepheid-hosting clusters given that the cluster stars within $\sim 1.5$\,pc appear to be bluer than the stars farther away from the center (cf. yellow upward triangles in Fig.\,\ref{fig:MeanBiasPerPc}). $\Delta m_{\rm{W,HVI}}$, which is particularly relevant for the measurement of $H_0$ is reduced compared to the bias measured using F160W exclusively. For a typical (observed M31) cluster Cepheid, we find a brightening by $\langle \Delta m_{\rm{W,HVI}} \rangle = -0.30 \pm 0.18$\,mag due to the integrated cluster population. If F160W observations are used without a color term, the typical bias is $\langle \Delta m_{\rm{F160W}} \rangle = -0.39 \pm 0.16$\,mag.

\begin{table*}
\centering
\begin{tabular}{lrrrrrrrrrr}
\hline
\hline
M31CC ID & $\Delta m_{\rm{F475W}}$ & $\Delta m_{\rm{F814W}}$ & $\Delta m_{\rm{F160W}}$ & $\Delta m_{\rm{VI}}$ & $\Delta m_{\rm{W,VI}}$ & $\Delta m_{\rm{W,HVI}}$ \\
 & (mag) & (mag) & (mag) & (mag) & (mag) & (mag) \\ 
 \hline
M31CC-01 & -1.7847 & -1.2192 & -1.0401 & -0.3929 & -0.6102 & -0.8829 \\
M31CC-02 & -0.2671 & -0.0400 & 0.1732 & -0.1696 & 0.2229 & 0.2410 \\
M31CC-03 & -0.9905 & -0.8469 & -0.5631 & -0.1120 & -0.6732 & -0.5183 \\
M31CC-04 & -0.5931 & 0.2327 & 0.4085 & -0.5624 & 1.1044 & 0.6335 \\
M31CC-05 & -0.2460 & -0.6125 & -0.8206 & 0.2150 & -0.9457 & -0.9066 \\
M31CC-06 & -0.5973 & -0.7054 & -0.9231 & 0.0714 & -0.8160 & -0.9516 \\
M31CC-07 & -0.8209 & -0.4311 & -0.2697 & -0.2787 & 0.0010 & -0.1582 \\
M31CC-08 & -0.7188 & -0.2448 & -0.4225 & -0.3349 & 0.2742 & -0.2885 \\
M31CC-09 & -1.0808 & -0.1819 & -0.0694 & -0.6094 & 0.7627 & 0.1744 \\
\hline
\hline
mean bias & -0.7888 & -0.4499 & -0.3918 & -0.2415 & -0.0756 & -0.2952  \\
standard mean error & 0.1476 & 0.1401 & 0.1569 & 0.0860 & 0.2294 & 0.1788  \\
median bias & -0.7188 & -0.4311 & -0.4225 & -0.2787 & 0.0010 & -0.2885  \\
\hline
intensity mean bias$^\ddagger$ & -0.9116 & -0.5611 & -0.5357 & -0.2542 & -0.1670 & -0.4340  \\
intensity mean error$^\ddagger$ & 0.1834 & 0.1638 & 0.1638 & 0.1648 & 0.3034 & 0.1766  \\
intensity median bias$^\ddagger$ & -0.6954 & -0.4359 & -0.4724 & -0.1924 & -0.1377 & -0.3954  \\
\hline
\hline
\end{tabular}
\caption{Cluster light contribution for 9 cluster Cepheids in M31, cf.
\S\ref{sec:M31ClusterCepheidPhotometry}.
Negative $\Delta m$ indicates brightening. M31CC ID is the
object number assigned in this paper, cf. Tab.\,\ref{tab:CCsample}.
$\Delta m_{\rm{VI}} = \Delta m_{\rm{F555W}} - \Delta m_{\rm{F814W}}$ is estimated
by interpolating the observed $\Delta m_{\rm{F475W}} - \Delta m_{\rm{F814W}}$ color in
color-color space (F475W-F814W vs F555W-F814W) defined by a 
PARSEC isochrone with $\mathrm{age} = 50\,$Myr and Solar metallicity.
$\Delta m_{\rm{W,VI}} = \Delta m_{\rm{F814W}} - 1.55\cdot \Delta m_{\rm{VI}}$ and $\Delta m_{\rm{W,HVI}} = \Delta m_{\rm{F160W}} - 0.4\cdot \Delta m_{\rm{VI}}$ quantify the bias in the 
reddening-free Wesenheit magnitudes.
We use mean and median biases computed column-wise for
all 9 Cepheids in magnitude space to estimate stellar association bias. Average biases computed in flux space are provided below for reference and marked by $^\ddagger$. 
\S\ref{sec:disc:averagebias} discusses the difference between the average bias estimates calculated in magnitude and flux space and explains our reasoning for adopting magnitude means for estimating stellar association bias.}
\label{tab:M31deltamag}
\end{table*}

\subsection{Cluster bias and $H_0$}
\label{sec:CC-H0}

We estimate the impact of stellar association bias due to cluster populations on $H_0$ by separately considering the frequency of Cepheids occurring in clusters and the average bias that applies to such Cepheids. The bias is computed in distance modulus by multiplying the M31 clustered Cepheid fraction, $f_{\rm{CC,M31}}$, with the average  bias measured on the Cepheid host clusters in M31, $\langle \Delta m_{\rm{CC,M31}} \rangle$: 
\begin{equation}
\Delta \mu_{\lambda}  = f_{\rm{CC,M31}} \cdot \langle \Delta m_{\rm{CC,M31}} \rangle_\lambda \ .
\label{eq:clusterbias_tot}
\end{equation}
Using the average biases listed in Tab.\,\ref{tab:M31deltamag} we find: $\Delta \mu_{\rm{WHVI}} = -7.4$\,mmag, $\Delta \mu_{\rm{H}} = -9.8$\,mmag, and $\Delta \mu_{\rm{WVI}} = -1.9$\,mmag.

We stress that the value of $f_{\rm{CC}}$ to be used for the above estimation must be representative of the \emph{effective} clustered Cepheid fraction ($f_{\rm{CC,eff}} < f_{\rm{CC,true}}$) as it applies to SN host galaxies across the cosmic distance ladder, taking into account selection effects and variations among galaxies as well as across different regions in external galaxies. We discuss these points in detail below (\S\ref{sec:disc:clusters}) and show that our estimation of $\Delta \mu_{\rm{CC}}$ based on M31 data actually provides an upper limit to stellar association bias on $H_0$, despite $f_{\rm{CC,M31}}$ being lower than the \emph{true} fractions estimated above for the MW, LMC, and SMC in \S\ref{sec:CCfrequency}.

\begin{figure}
\centering
\includegraphics{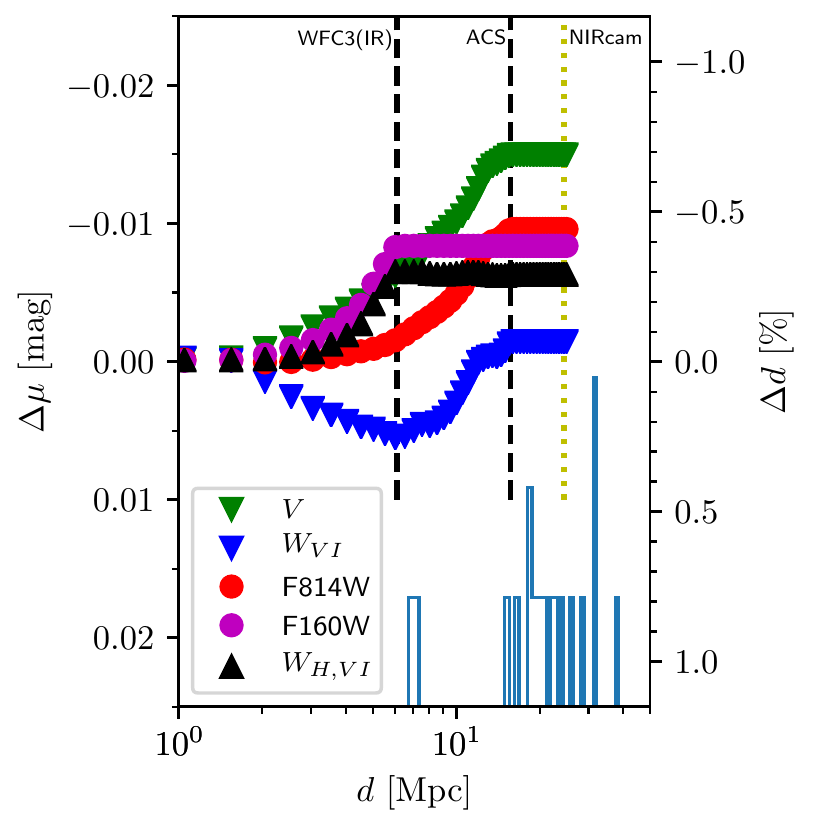}
\caption{Combined stellar association distance bias due to both binaries and clusters ($\Delta \mu$ or $\Delta d$) incurred within a single pixel as a function of the distance at which a galaxy of interest is located. The curves level off at the distance where the typical physical cluster size ($3.8$\,pc, cf. \S\ref{sec:CCPhotBias}) is equal to the size of a single pixel given the WFC(IR) and ACS plate scales (cf. vertical dashed lines). The dotted yellow line indicates the distance at which one JWST/NIRcam pixel corresponds to a physical scale of $3.8$\,pc. The histogram at the bottom shows the distribution of {\it SH0ES} SNIa-host galaxies (R+16). M101 is the sole SN-host galaxy for which resolving Cepheid host clusters may be feasible. The positive $\Delta \mu$ in the $W_{VI}$ filter combination is a consequence of the typical cluster color and the definition of this Wesenheit magnitude. Interestingly, the total stellar association bias is very near $0$ in the optical ($V-I$) Wesenheit magnitude.}
\label{fig:BiasVSDistance}
\end{figure}

Propagating this bias to the determination of $H_0$, it must be kept in mind that stellar association bias arises due to the distance-dependent ability to spatially resolve clusters, cf. Fig.\,\ref{fig:BiasVSDistance}. 
Star clusters are barely resolved at the distance of NGC\,4258 ($7.5$\,Mpc) or M101 ($6.7$\,Mpc, R+16), since typical cluster half-light radii of 2 pc correspond approximately to the plate scale of {\it HST/ACS}. Since NGC\,4258 is similarly affected by cluster light contributions as the 19 {\it SH0ES} SN-host galaxies, the stellar association bias does not apply in this galaxy, especially when working with F160W magnitudes. However, for the two other anchors used to set up the distance ladder (MW and LMC), this bias does apply. Adopting the three anchors recommended by R+16 (MW, LMC, NGC\,4258) and weighting them equally\footnote{We have verified that this is a good approximation by running the global optimization as described in R+16}, it follows that the impact of stellar association bias should be weighted by two thirds.

For the distance scale as a whole, we thus estimate the effective bias in distance modulus to be:
\begin{equation}
\Delta \mu_{\rm{F160W}} = 2/3 \cdot \Delta \mu_{\rm{CC,F160W}} = -6.5\,\rm{mmag} 
\end{equation}
\begin{equation}
\Delta \mu_{\rm{W_{HVI}}} = 2/3 \cdot \Delta \mu_{\rm{CC,WHVI}} = -4.9\,\rm{mmag}\ .
\end{equation}

These distance modulus differentials translate to fractional distance bias of 
$0.30\%$ and $0.23\%$ in F160W and $W_{\rm{H,VI}}$, respectively. This would imply the R+16 measurement of the Hubble constant ($H_0 = 73.24$\,km\,s$^{-1}$\,Mpc$^{-1}$) to be overestimated by $\Delta H_0 = 0.22$ and $0.17$\,km\,s$^{-1}$\,Mpc$^{-1}$. However, we stress that this bias is an upper limit to the true bias incurred, since Cepheids heavily affected by significant cluster populations are precluded from being a part of the analysis due to discovery bias, cf. \S\ref{sec:disc:clusters}. Stellar association bias due to cluster populations is thus at more than an order of magnitude larger than the bias due to wide binaries (\S\ref{sec:binaries}), because clusters contain hundreds to thousands of stars, which of course by far outweighs the fact that the occurrence rate of Cepheids in clusters is $\sim 3$ times lower than the occurrence of wide binaries.

Correcting for stellar association bias thus leads to $H_0 = 73.07 \pm 1.76\, \rm{km\,s^{-1}\,Mpc^{-1}}$ (using the $W_{\rm{H,VI}}$ bias estimate), which remains (nearly unchanged) $3.3\sigma$ larger than $H_0$ determined by {\it Planck} \citep[cf. tab.\,8 in][$66.93 \pm 0.62\, \rm{km\,s^{-1}\,Mpc^{-1}}$]{2016A&A...596A.107P}. Thus, correcting for stellar association bias does not reconcile the observed tension between the distance scale and CMB-based values of $H_0$.

\section{Discussion}\label{sec:discussion}

\subsection{Cepheids in binary systems and the distance scale}
\label{sec:disc:binaries}

Binary companions affect the PLR calibration in at least two ways: one, by providing additional flux and two, by biasing parallax measurements for Galactic Cepheids, if orbital motion is not taken into account in the astrometric model. Companions with reasonably short orbital periods (a couple to a few years)
capable of significantly affecting parallax are detectable 
using precision radial velocity measurements \citep{2016ApJS..226...18A}, even if such companion stars are too faint to leave a clear photometric signature \citep{2015ApJ...804..144A,2016MNRAS.461.1451G}. In most such cases, {\it Gaia} \citep{2016A&A...595A...1G,2017A&A...605A..79G} astrometry should be sufficient to solve for orbital motion, in particular when independent evidence of a companion's presence is available a priori. 

In general, the ``parasitic'' flux contributed by companion stars cannot be corrected individually, although this may be possible statistically. As shown in \S\ref{sec:binaries}, additional flux contributed by wide binaries has a negligible effect on the distance scale, in particular when working with near-IR data as well as Wesenheit magnitudes (thanks to the color difference between Cepheids and typical companions). However, a larger bias for the distance scale would arise if the Galactic PLR calibration were based exclusively on single Cepheids, while extragalactic Cepheids were not selected accordingly. Bias due to binaries would increase in such a case primarily because $f_{\rm{wb}}$ would have to be replaced with the total binary fraction of Cepheids. Not correcting for differences in preferred mass ratios in closer-in binary systems \citep{2017ApJS..230...15M}, Tab.\,\ref{tab:magnitudebias} provides a useful upper limit to this effect, which would be approximately $\Delta m_{W_{\rm{H,VI}}} = 0.6 \cdot -1.7 = -1.0$\,mmag, i.e., $-0.05\%$ in distance for 20\,d Cepheids.

Our results suggest a period-dependent contrast between Cepheids and their companions, which leads to a predicted, albeit small, non-linearity in the PLR slope of real Cepheid populations consisting of both singles and multiples. Using the numbers listed Tab.\,\ref{tab:magnitudebias} and assuming a total binary fraction of $f_{\rm{bin}} = 60\%$, we find a period-dependent deviation (computed as the difference between $5$\,d and $40$\,d Cepheids) from the single-star PLR of $-13$\,mmag in $V$-band, whereas the deviation reduces to $-7.8$, $-4.6$, $-3.1$, and $-3.5$\,mmag in $I$- and $H$-band and the Wesenheit magnitudes $W_{\rm{VI}}$ and $W_{\rm{H,VI}}$. These numbers underline that non-linearity due to binaries is a small effect, even if they are upper limits, since closer-in binaries favor smaller mass ratios than wide binaries \citep{2017ApJS..230...15M}. Observational evidence for PLR non-linearity has been disputed in the literature, indicating a weak phenomenon. However, PLR non-linearity appears to be more readily observed in optical passbands than in near-IR or Wesenheit magnitudes \citep{2002Ap&SS.280..165T,2004A&A...424...43S,2006ApJ...650..180N,2013ApJ...764...84I,2016MNRAS.457.1644B}. These two facts seem to corroborate an origin of PLR non-linearity linked to multiplicity, since the contrast between Cepheids and their companions depend on wavelength and are significantly smaller in $V$-band than in $H$-band. However, given the disputed nature of PLR non-linearity in the literature, it is clear that further study is required to conclude in this matter.

\subsection{Cepheids in clusters and the distance scale}
\label{sec:disc:clusters}

The estimation in \S\ref{sec:CC-H0} represents an upper limit of the actual degree to which the cosmic distance scale and $H_0$ are actually biased by Cepheids occurring in clusters. To elucidate this subtle and important point, this section discusses several effects that impact both the \emph{effective} clustered Cepheid fraction, $f_{\rm{CC,eff}}$, and the \emph{effective} average bias due to clusters, $\Delta \mu_{\rm{CC,eff}}$. 
Additional observations and further investigations are required to quantify these effects in more detail.

\label{sec:disc:discoverybias}
Firstly, \emph{discovery bias} reduces $f_{\rm{CC,eff}}$ relative to the \emph{true} $f_{\rm{CC}}$ in distant galaxies, since strict selection criteria applied to extragalactic Cepheid candidates \citep[cf.][]{2011ApJ...730..119R,2013AJ....145..106K,2015MNRAS.451..724W} preclude heavily blended Cepheids from being used in the distance scale calibration. \citet[their Sec.\,4.2]{2016ApJ...830...10H} describe the selection criteria applied to Cepheids in the {\it SH0ES} project in detail. The criteria include light curve shape (fit agreement with template light curves and visual inspection), amplitudes, mean colors, and proximity to the Leavitt Law in a given galaxy. A period dependence of discovery bias is expected, because of lower intrinsic amplitudes among Cepheids with $P_{\rm{p}} \sim 10$\,d and $> 40$\,d. Quantifying this discovery bias including its dependence on \Ppuls, (unbiased) amplitude, reddening, and galaxy-specific aspects such as metallicity or background levels, etc., is not straightforward, however, and outside the scope of this work.

In a previous investigation of blending-related issues on the distance scale, \citet{2000PASP..112..177F} noted that contamination levels above $60\%$ ($-0.5$\,mag) would usually be discernible during light curve inspection, whereas contamination levels below ($30\%$) would usually not be noticed. 
We measure an average bias of $-0.79$\,mag in F475W, cf. Tab.\,\ref{tab:M31deltamag}, ($-0.70$\,mag in F555W), which implies clearly noticeable effects on Cepheid light curves in short-wavelength optical bands. This notably includes F350LP, which is used for Cepheid discovery (effective wavelengths: $0.53\,\mu$m and $0.61\,\mu$m for F555W and F350LP). Hence, the cluster Cepheids studied here would most likely not form part of the observed Cepheid Leavitt Law in more distant galaxies.

\begin{figure}
\centering
\includegraphics{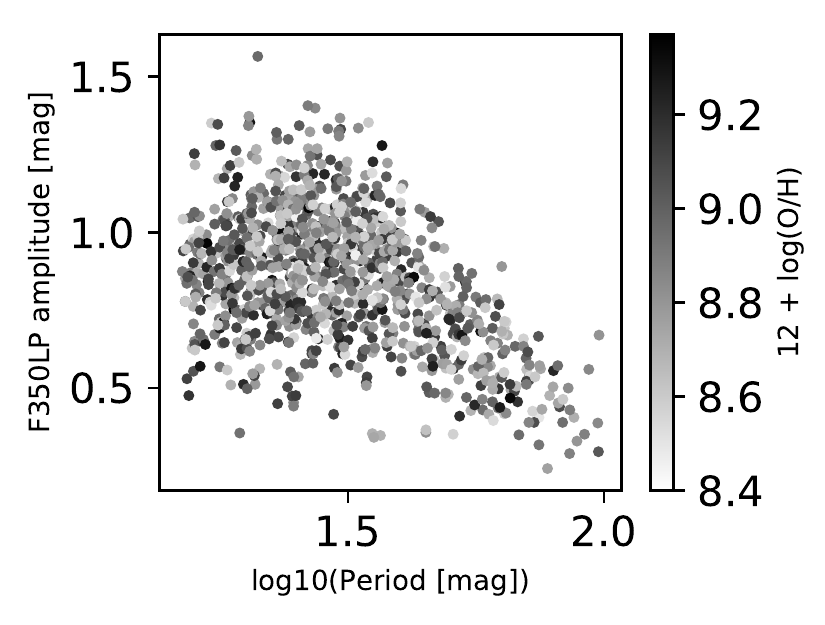}
\caption{F350LP amplitudes of Cepheids in SN host galaxies against period and oxygen abundance \citep[based on Tab.\,5 in][]{2016ApJ...830...10H}. \label{fig:disc:amplitudes}}
\end{figure}

Figure~\ref{fig:disc:amplitudes} confirms that Cepheid samples in SN host galaxies are effectively cleaned of heavily blended objects. The F350LP amplitudes are not significantly depressed compared to $V-$Band amplitudes of MW Cepheids \citep[their Fig.\,1]{2009A&A...504..959K}. We stress, however, that R+16 do correct for the effect of blended background stars on mean magnitudes using artificial star tests.

Secondly, our bias estimation in \S\ref{sec:CC-H0} provides an \emph{upper limit on stellar association bias}, because M31 clusters are less luminous and less numerous than clusters in other galaxies \citep[their Fig.\,17]{2015ApJ...802..127J}. This is based on two pieces of information. On the one hand, cluster formation efficiency, $\Gamma$, depends on star formation surface density $\Sigma_{\rm{SFR}}$ \citep{2016ApJ...827...33J,2017ApJ...839...78J}, and this effect can likely explain the lower $f_{\rm{CC,M31}}$ value compared to other galaxies (cf. \S\ref{sec:CCfrequency}). 
On the other hand, (Cepheid host) clusters in other galaxies tend to be more luminous than in M31 and virtually no low-luminosity clusters exist in galaxies with higher $\Sigma_{\rm{SFR}}$. Therefore, the aforementioned discovery bias is typically even more effective at removing cluster Cepheids in other galaxies, and more luminous clusters than the ones presented her would be precluded from affecting the distance scale.

Different regions within a given galaxy are expected to have different \emph{true} values of $f_{\rm{CC}}$ and different cluster luminosity functions, since $\Sigma_{\rm{SFR}}$ varies across galaxies, e.g. with galactocentric radius \citep{2015MNRAS.452..246A,2016ApJ...827...33J}. As Fig.\,3 in R+16 shows, most Cepheids in SN host galaxies are  discovered far from the centers of their host galaxies, where $\Sigma_{\rm{SFR}}$ reaches values more comparable to M31, cf. \citet[their Tab.\,2]{2015MNRAS.452..246A} and \citet{2016ApJ...827...33J}.

Finally, we note that occasional chance blending by clusters during artificial star tests does not significantly compensate for stellar association bias. Based on a cross-match (cf. Fig.\,\ref{fig:ClnearCepM31}) of AP-clusters located within $6.3$\,arcmin (corresponds to 100 {\it WFC3/IR} pixels at the mean {\it SH0ES} SN-host distance) of PS1 Cepheids, we compute a fractional area\footnote{Computed using twice the effective AP cluster radii} occupied by clusters of $2.2 \times 10^{-5}$, two orders of magnitude smaller than $f_{\rm{CC}}$.

In summary, M31 provides the best available upper limit on $f_{\rm{CC}}$ and the average cluster bias. More luminous Cepheid host clusters in other galaxies are even more likely to be filtered out by Cepheid discovery bias.
Future studies related to cluster mass functions of galaxies should provide further guidance. In particular M101 will provide useful information in this regard, since the prospects for spatially resolving clusters using {\it JWST} are the best in this {\it SH0ES} SN host galaxy.

\subsubsection{Averaging cluster bias in magnitude and flux space\label{sec:disc:averagebias}}

Following a comment from the referee, we have calculated the typical cluster bias in intensity space as well as in magnitude space. Averaging the contributions from the 9 clusters yields a noticeably different typical bias estimate, cf. Tab.\,\ref{tab:M31deltamag}. However, working with flux-averages implies a different interpretation of this bias due to different underlying assumptions. Specifically, averaging in magnitude space implies that the bias is \emph{multiplicative in flux space}.
Averaging in flux space implies that the bias is \emph{additive in flux space}. 
As a result, the estimation of the total bias as it applies to $H_0$ must take into account the average luminosity of Cepheids in SN host galaxies when working with intensity averages.

Cepheids in SN host galaxies are on average more luminous than M31 cluster Cepheids. To wit, $\langle \log{P_{\rm{p,CC,M31}}} \rangle \approx 1.08$ among M31 cluster Cepheids and $\langle \log{P_{\rm{p,SN}}} \rangle \approx 1.45$ among Cepheids in SN-host galaxies (tab.\,4 in R+16, not counting Cepheids in the anchor galaxies M31 and NGC\,4258, since M31 is excluded from the R+16 recommended solutions and clusters are not spatially resolved in NGC\,4258).
Hence, the intensity-averaged distance modulus bias is rescaled as:
\begin{equation}
\Delta \mu_{\rm{CC},\log{P_{\rm{p}}}} = -2.5 \log{\left( 1 + \frac{10^{-0.4 \Delta \mu_{\rm{CC,M31}} } - 1}{ 10^{-0.4 b \cdot \delta\log{P_{\rm{p}}}}} \right)} \ ,
\label{eq:clusterbias_logPcorr}
\end{equation}
where $b = -3.27$ ($-3.06$) is the slope of the PLR in $W_{\rm{H,VI}}$ (F160W) magnitudes (R+16, their Tab.\,8) and $\delta\log{P_{\rm{p}}} = 1.45 - 1.08 = 0.37$.
Hence, the flux-averaged $\Delta \mu_\lambda$ is corrected downward by a factor of $3.0$ ($2.8$), yielding $\Delta \mu_{\rm{CC},\log{P_{\rm{p}}}} = -3.8$\,mmag ($-4.7$\,mmag) and an $H_0$ bias of $0.15\,\%$ ($0.12\,\%$) using $W_{\rm{H,VI}}$ (F160W), which is about half the impact estimated in \S\ref{sec:CC-H0}.

We presently do not have sufficient information to understand whether cluster bias is more constant in magnitude or in flux space. However, we have a small preference for considering the bias constant in magnitude, since there are physical reasons to expect host populations of longer-\Ppuls\ cluster Cepheid to be brighter than those of their shorter-period counterparts. Specifically, Cepheid masses correlate with \Ppuls, and total cluster mass depends the most massive member's mass \citep{2010MNRAS.401..275W}, and more massive clusters are more luminous \citep[for MW clusters, cf.][]{2011A&A...525A.122P}. 
However, turning cluster mass into light is inefficient and the contrast between an evolved Cepheid and the hottest MS cluster members increases with \Ppuls, in particular when working with Wesenheit or infrared magnitudes. 
Thus, the selection bias towards more long-period Cepheids in extragalactic samples  decreases average cluster bias, even when working with magnitude averages.

\subsection{Considerations regarding $f_{\rm{CC,true}}$ \label{sec:disc:fcc}}

The \emph{effective} clustered Cepheid fraction as it applies to Cepheids in SN host galaxies linearly affects the result of the total bias (cf. \S\ref{sec:CC-H0} and \ref{sec:disc:clusters}), and is thus one of the key uncertainties for this work. However, the \emph{true} $f_{\rm{CC}}$ provides interesting information on several astrophysical processes and deserves further attention. Here, we discuss the completeness of the estimation of \emph{true} value of $f_{\rm{CC,M31}}$ (\S\ref{sec:disc:fcc:completeness}) and a possible dependence of $f_{\rm{CC,true}}$ on age (\S\ref{sec:disc:fcc:age+gal})\hbox{---}and thus \Ppuls.

\begin{figure}
\centering
\includegraphics{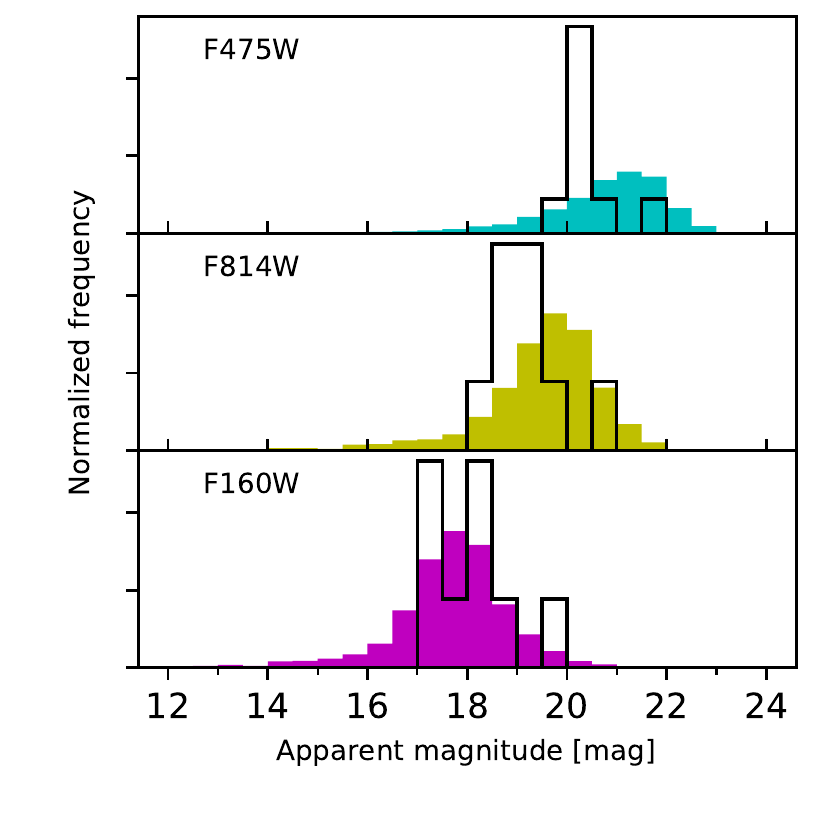}
\caption{Luminosities of all PHAT clusters \citep[filled in
step historgrams]{2015ApJ...802..127J} compared to the luminosities of 9
Cepheid-hosting clusters in M31 (black step histograms). All histograms are normalized to unit area.}
\label{fig:HistoClusterLumis}
\end{figure}

\subsubsection{Completeness of $f_{\rm{CC,M31}}$ \label{sec:disc:fcc:completeness}}

Figure\,\ref{fig:HistoClusterLumis} shows normalized distributions of AP cluster integrated magnitudes and highlights known Cepheid host clusters for optical and near-IR passbands. No selection based on age was made when preparing these histograms. The figure illustrates the result of \S\ref{sec:CCPhotBias} that Cepheid host clusters have fairly hot member stars, i.e., have blue integrated colors, cf. Fig.\,\ref{fig:CCpostagestamps}. The blue intrinsic color of Cepheid host clusters may thus be exploited to identify extragalactic Cepheids that are blended with host clusters. UV and short wavelength optical ($U$ or $B$-band) observations of Cepheids in {\it SH0ES} galaxies would be useful for identifying such cluster contamination.

To ensure an unbiased estimate of $f_{\rm{CC,M31}}$, we decided to exploit the blue intrinsic color of host clusters (cf. Figs.\,\ref{fig:UVclusterdetectability} and \ref{fig:HistoClusterLumis}) and visually inspected UV (F275W and F336W filters) postage stamps similar to Fig.\,\ref{fig:CCpostagestamps} for all 453 FM or UN PS1 Cepheids for the presence of any additional clusters that were not identified by the AP. These UV postage stamps are included in the online appendix \ref{app:postagestamps}. For comparison, Fig.\,\ref{app:fig:UVPS_CCs} highlights the previously known M31 cluster Cepheids \citep{2015ApJ...813...31S}.

The inspection of UV postage stamps reveals the AP cluster catalog to be highly complete in terms of detecting significant\footnote{containing, say, more than 4 blue stars in F275W and F336W} clusters near Cepheids. 
We found no obviously significant clusters located near any of the 442 Cepheids not already known to be cluster members. However, we have highlighted 17 Cepheids in in Fig.\,\ref{app:fig:UVPS_interesting} near which the density of blue stars appears to be higher than near most other Cepheids. Specifically, Fig.\,\ref{app:fig:UVPS_interesting} includes J011.5948+41.9405, which may feature a cluster off-center from the Cepheid, similar to the known cluster Cepheid J011.0209+41.3162. Two further cases, J011.6531+42.2074 and J011.2147+41.8490, may qualify as clusters or diffuse groups, depending on what definition to adopt for this terminology. However, these groupings are clearly different from the significant Cepheid host clusters shown in Fig.\,\ref{app:fig:UVPS_CCs}.
Conversely, the reality of two previously identified Cepheid-hosting clusters appears questionable: J011.6519+42.1286 (M31CC-05) and J011.7119+42.0449 (M31CC-06). Moreover, the curves of growth near these two Cepheids (Fig.\,\ref{fig:COGcluster}) reveal very red populations, which suggests erroneous identifications as clusters due to field star contamination.

As the inspection of the UV postage stamps shows, the definition of the term `star cluster' may be insufficiently well defined to objectively discriminate between obvious clusters and other, more diffuse, types of physical stellar associations. However, there is a clear absence of intermediate objects and the contribution to $H_0$ by any diffuse groups such as near J011.6531+42.2074 and J011.2147+41.8490 would be negligible due to the high color contrast between Cepheids and young stars and the use of near infrared observations and Wesenheit magnitudes. We plan to extend the present investigation to include any types of stellar association in the future, regardless of definitions such as `cluster'.

A further interesting point to consider is that $f_{\rm{CC,true,M31}} = 2.5\%$ is lower than the clustering efficiency suggests $\Gamma_{\rm{M31}} = 4 - 8\%$ \citep{2015ApJ...802..127J}. 
Binary interactions and mergers can significantly reduce the number of stars that survive until the Cepheid evolutionary stage, and this fraction has been estimated to be $\mathcal{F}_{\rm Cepheid} = 0.75 \pm 0.15$ for mid-B stars \citet[their Sect.\,6.1]{2017ApJS..230...15M}. Considering on $f_{\rm{CC,M31}} / \Gamma_{\rm{M31}} = 0.31 - 0.62$, we find tentative signs of the survival rate being lower in clusters, where additional dynamical interactions and mergers may occur.

\subsubsection{Dependence of $f_{\rm{CC}}$ on host galaxy and time \label{sec:disc:fcc:age+gal}}

As argued in \S\ref{sec:disc:clusters}, there are physical reasons to expect variations in the true fraction of clustered Cepheids on the galaxy considered, or even the position therein. Additionally, cluster dispersal timescales ($\sim 10$\,Myr) are shorter than, yet relatively close to, the ages of Cepheid ($\sim 30 - 300$\,Myr), so that a trend with Cepheid age can be expected. The clearest example of such differences between galaxies is the LMC that contains clusters such as NGC\,1866 with more than $20$ (short-period) Cepheids for which there are no known analogs in any other galaxy. Of course, the various estimates of $f_{\rm{CC}}$ presented in \S\ref{sec:CCfrequency} are subject to a complex mixture of completeness and contamination issues and selection effects, which depend on the galaxy being considered. We plan to further investigate  these matters using {\it Gaia} data and Nbody simulations.

\begin{figure*}
\centering
\includegraphics{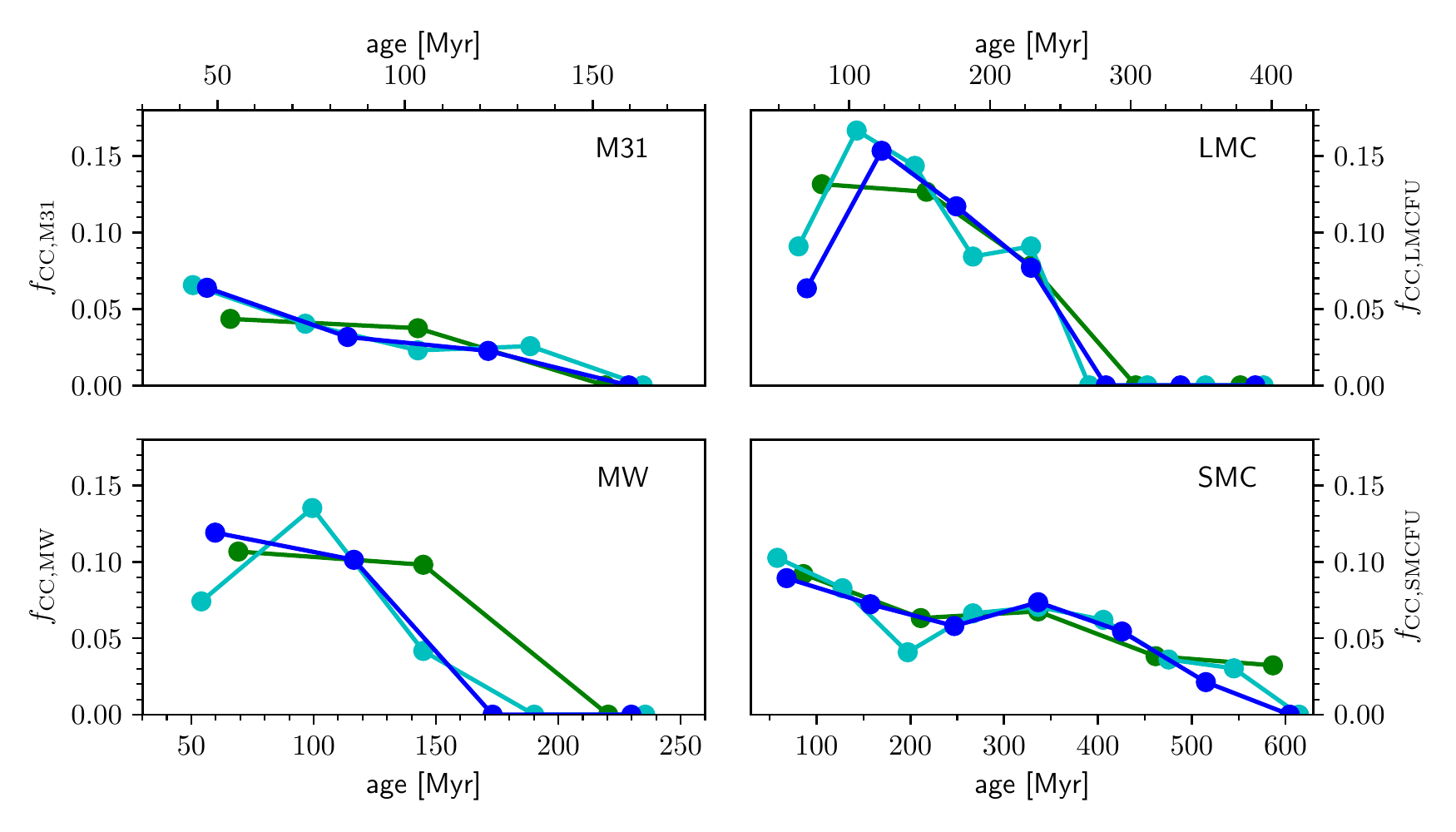}
\caption{Dependence of the observed $f_{\rm{CC}}$ on galaxy and age. Ages are calculated using period-age relations \citep{2016A&A...591A...8A} for Solar metallicity ($Z=0.014$; MW, M31) and metallicities corresponding to the LMC ($Z=0.006$) and the SMC ($Z=0.002$). The fractions are computed using the PHAT footprint for M31, Cepheids within $2$\,kpc from the Sun for the Milky Way, and the OGLE-II catalogs of Cepheids and cluster Cepheids (multiple mentions removed), cf. \S\ref{sec:CCfrequency}. Three curves per panel are shown (MW and M31: 3, 4, and 5 bins; LMC and SMC 5, 7, and 9 bins shown in green, blue, and cyan, respectively) to illustrate the sensitivity of these trends on binning due to small number statistics. \label{fig:clusteragefcc}}
\end{figure*}

For the time being, we find tentative evidence of an age (period) dependence of $f_{\rm{CC,true}}$ in all galaxies considered, including M31, the MW, the LMC, and the SMC, cf. Fig.\,\ref{fig:clusteragefcc}. 
We speculate that this age trend traces the continued dispersal of clusters over the age range spread by Cepheids. 
Since \Ppuls\ correlates with age and IS position, $f_{\rm{CC,true}}$ is expected to have a period-(color) dependence, which could in principle affect the slope of the Leavitt Law and lead to PLR non-linearity. Specifically, longer-period Cepheids would be biased bright compared to shorter-period Cepheids, increasing the slope of the PLR. However, we stress that this is a second order effect to an already small bias and at present we do not have the statistics to test this in detail.

\subsection{Mitigating Stellar Association Bias}
\label{sec:avoidingBias}

Thanks to the upcoming releases of {\it Gaia} parallaxes, the calibration of the Leavitt law in the MW will experience a quantum leap in accuracy, and harnessing {\it Gaia}'s power to push toward an $H_0$ measurement to within $1\%$ accuracy requires mitigation of any intervening uncertainties and biases. Here, we have estimated that stellar association bias amounts to approximately $0.2\%$ of $H_0$. Hence, stellar association bias shall not prevent a measurement of $H_0$ to within $1\%$, despite the stated uncertainty involving the effective fraction of Cepheids occurring in clusters. Nonetheless, we identify a few strategies for avoiding or mitigating stellar association bias below. 

\begin{enumerate}
\item MW Cepheids in binary (multiple) systems should not be excluded from local PLR calibrations, since no analogous selection is made in the case of extragalactic Cepheids for which no knowledge regarding their multiplicity is available. This is equivalent to assuming a universal binary fraction for Cepheids. Calibrating the MW PLR exclusively using single Cepheids implies a passband-dependent slope difference between the MW and extragalactic Leavitt laws, cf. \S\ref{sec:disc:binaries}.
\item Additional independent distance estimates to SN-host galaxies would be extremely useful, provided they are insensitive to blending. This is the case for parallax measurements of the water mega-maser in NGC\,4258 \citep{2013ApJ...775...13H}, for instance. It remains to be seen how many more such maser distances can be determined. 
\item Observations in UV and short wavelength optical passbands may be useful for identifying Cepheids residing in comparatively young clusters via differences in average colors. Stellar association bias would be very effectively mitigated, if cluster Cepheids could thus be identified and excluded from the analysis.
\item UV data of the SN-host galaxy M101 and the anchor galaxy NGC\,4258 would be particularly useful to obtain a second distance-scale relevant estimate the effective fraction of clustered Cepheids, $f_{\rm{CC}}$, cf. Fig.\,\ref{fig:BiasVSDistance}. 
\item The improved spatial resolution of {\it JWST}'s NIRCAM  in combination with existing optical {\it HST} imaging may allow to estimate the average $W_{\rm{H,VI}}$ bias in M101.
\item Stellar association bias is expected to be reduced for long-period Cepheids than for short-period Cepheids given the increased luminosity contrast between a Cepheid and its host cluster population, despite long-period Cepheids likely occurring more frequently in clusters (cf. Fig.\,\ref{fig:clusteragefcc}).
\end{enumerate}

\section{Summary}\label{sec:summary}

We have investigated the effect of stellar association bias on Cepheid distance measurements and the Hubble constant. 
Understanding this and other possible biases affecting the distance scale is crucial for measuring $H_0$ with an accuracy of $1\%$ to elucidate dark energy and understand the origin of the observed {\it tension} between distance scale and CMB-based determinations of $H_0$.

We have estimated stellar association bias in two parts. First, we have used state-of-the-art stellar evolution models \citep{2012A&A...537A.146E} and detailed statistics on the multiplicity fraction of intermediate-mass stars to investigate the effect of wide binaries. We show that wide binaries have a negligible effect of approximately $0.004\%$ on $H_0$. 

Second, we have used deep {\it HST} imaging of M31 provided by the PHAT survey \citep{2012ApJS..200...18D} to measure the photometric contribution by typical Cepheid hosting clusters in photometric passbands commonly used to construct the distance scale. We have further considered the clustered fraction of Cepheids, $f_{\rm{CC}}$ in different galaxies (MW, LMC, SMC, M31). We find that the true value of $f_{\rm{CC}}$ likely varies considerably among galaxies and within a given galaxy, as well as with time (age). However, discovery bias reduces the effective fraction applicable to the distance scale calibration, $f_{\rm{CC,eff}}$, considerably, thus implying a small overall effect of stellar association bias on the distance scale. We find the M31 based estimate of $f_{\rm{CC,M31}} = 2.5\%$ to be the most adequate number for the estimation of stellar association bias due to clusters.   

Using M31's cluster Cepheid population, we estimate the $H_0$ bias due to star clusters at approximately $-0.30\%$ and $-0.23\%$, respectively using $H-$band (F160W) observations and Wesenheit magnitudes $W_{\rm{H,VI}}$ based on F160W, F555W, and F814W observations. We notice that this represents an upper limit to the likely bias level, since the present sample of M31 cluster Cepheids would have likely been precluded from further analysis due to discovery bias and sample selection.

We have thus shown that stellar association bias does not prevent achieving $1\%$ accuracy on $H_0$, and that stellar association bias does not explain recently discussed tension between $H_0$ values based on the extragalactic distance scale and the CMB. Further research into this area is warranted, in particular to better understand $f_{\rm{CC,eff}}$. Short optical wavelength observations of nearby Cepheid hosting galaxies would be particularly beneficial to this end.

\acknowledgments

We thank the anonymous referee for their detailed report and Stefano Casertano for useful discussions. We further acknowledge a useful comment by L.~Clifton Johnson related to M31's star formation surface density. 

Based on observations made with the NASA/ESA Hubble Space Telescope, obtained from the data archive at the Space Telescope Science Institute. STScI is operated by the Association of Universities for Research in Astronomy, Inc. under NASA contract NAS 5-26555.

This research made use of NASA's Astrophysics Data System Bibliographic Services and Astropy, a community-developed core Python package for Astronomy \citep{astropy}.

\vspace{5mm}
\facilities{HST(ACS and WFC3)}

\software{astropy \citep{astropy}}

\bibliographystyle{aasjournal}
\bibliography{Biblio}

\begin{appendix}
\label{sec:app}

\section{List of Milky Way Cepheids within $d < 2$\,kpc from the Sun \label{app:MWCeps}}
The following Cepheids, ordered by distance, were used to estimate $f_{\rm{CC,MW}}$ in \S\ref{sec:CCfrequency}: 
$\eta$~Aql, X~Sgr, $\beta$~Dor, $\zeta$~Gem, RT~Aur, W~Sgr, Y~Sgr, T~Vul, $\ell$~Car, AX~Cir, Y~Oph, U~Aql, V636~Cas, U~Sgr, U~Vul, R~TrA, S~Sge, V~Cen, S~Cru, AW~Per, RV~Sco, BF~Oph, T~Cru, V636~Sco, R~Cru, SU~Cyg, S~TrA, AP~Sgr, BB~Sgr, R~Mus, S~Mus, V737~Cen, RX~Cam, X~Vul, S~Nor, V350~Sgr, W~Gem, V482~Sco, AP~Pup, SS~Sct, TT~Aql, V~Vel, FM~Aql, V~Car, BG~Vel, T~Vel, V386~Cyg, ST~Tau, X~Cyg, V381~Cen, ER~Car, YZ~Sgr, XX~Sgr, RY~Sco, V912~Aql, V1162~Aql, ASAS~J124435-6331.8, AG~Cru, RY~CMa, TX~Cyg, FN~Aql, RT~Mus, T~Mon, U~Nor, RS~Cas, MS~Mus, V773~Sgr, V500~Sco, AS~Per, UX~Car, MW~Cyg, VW~Cru, SU~Cru, RZ~Vel, AU~Peg, RS~Ori, GSC~03996-00312, U~Car, V~Lac, AT~Pup, X~Cru, CR~Ser, V495~Cyg, V470~Sco, GU~Nor, RX~Aur, MY~Cen, V600~Aql, V412~Ser, XX~Cen, X~Sct, BE~Mon, ASAS~J175957-0348.7, BR~Vul, V514~Cyg, CV~Mon, AY~Cen, BG~Lac, DL~Cas, Y~Sct, AY~Sgr, SX~Car, V496~Cen, RR~Lac, RZ~CMa, VZ~Cyg, Y~Aur, UW~Car, IQ~Nor, V520~Cyg, Z~Lac, RW~Cam, VY~Car, VY~Cyg, V340~Nor, ASAS~J184443-0401.5, SV~Vul, RU~Sct, XY~Cas, SY~Cas, KX~Cyg, ASAS~J184741-0654.4, SY~Nor, V339~Cen, WZ~Sgr, RS~Pup, HO~Vul, SW~Cas, and SX~Vel. No further information is included at this time, since this list will soon be revised by {\it Gaia}.

\section{Postage Stamp Images of M31 Cepheids in {\it HST}'s UV passbands F275W and F336W\label{app:postagestamps}}

\begin{figure*}
\centering
\includegraphics{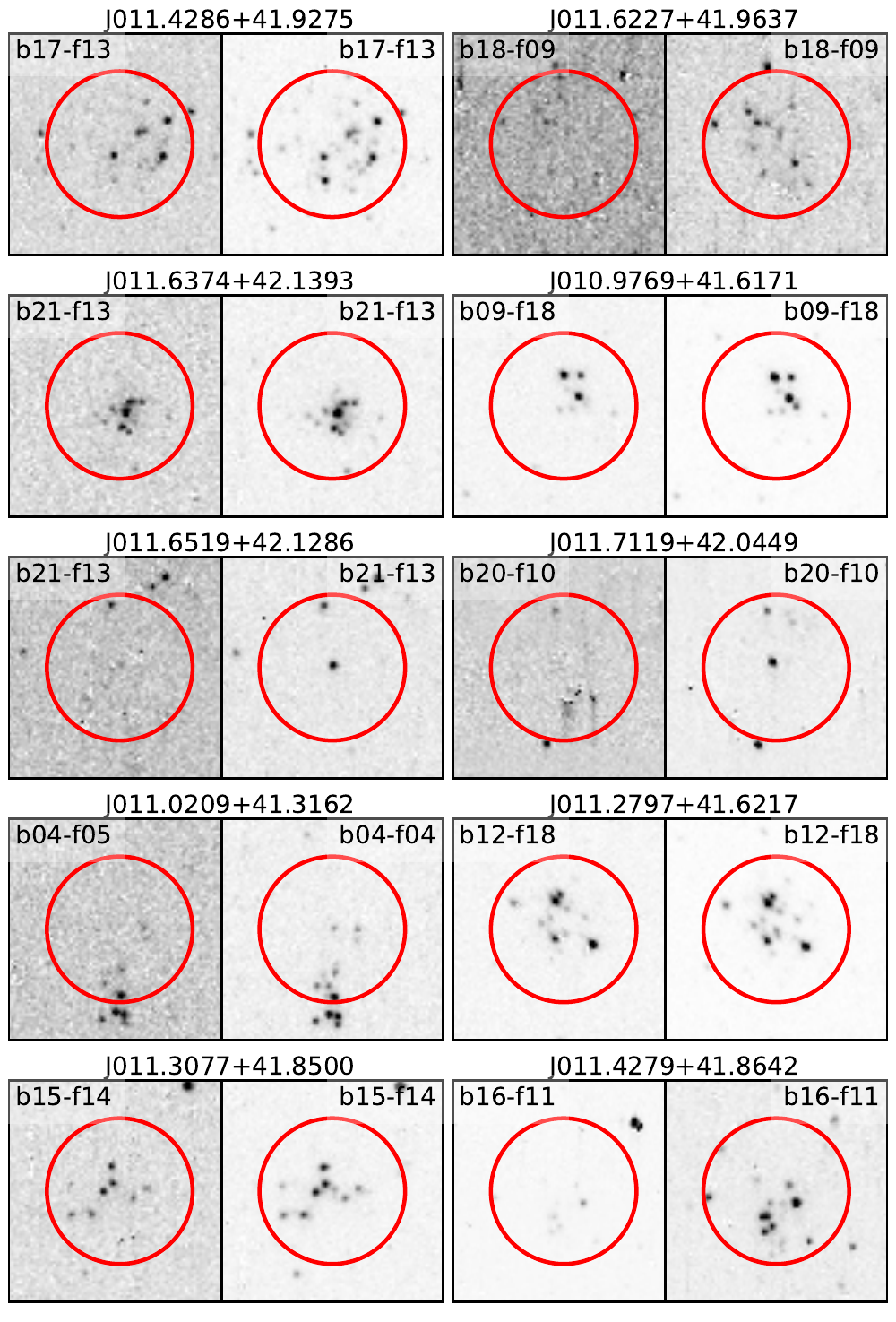}
\caption{Image cutouts (``postage stamps'') near M31 cluster Cepheids \citep{2015ApJ...813...31S} observed using {\it HST/WFC3}. Each postage stamp measures $3.2 \times 3.2\,\rm{arcsec}^2$ (80 x 80 pixel$^2$). All data shown were observed as part of the PHAT project \citep{2012ApJS..200...18D} and retrieved from MAST. For each star, we show an F275W observation in the left panel and an F336W observation in the right panel. The PHAT project field and brick numbers are indicated as b[brick number]-f[field number] in each panel. The red circle is centered on the (ground-based) PS1 Cepheid position and has a radius of $26.3$ pixel, which corresponds to a physical size of $3.8$\,pc at M31's distance of $774$\,kpc (R+16) to match the typical cluster size determined in \S\ref{sec:CCPhotBias}. To ensure a clear linear gray scale, we have limited the image to the $[0.02, 99.9]\%$ range of counts. Real clusters are expected to be dominated by numerous hot young main sequence stars that would be clearly visible in F275W and F336W.  \label{app:fig:UVPS_CCs}}
\end{figure*}

\begin{figure*}
\centering
\includegraphics{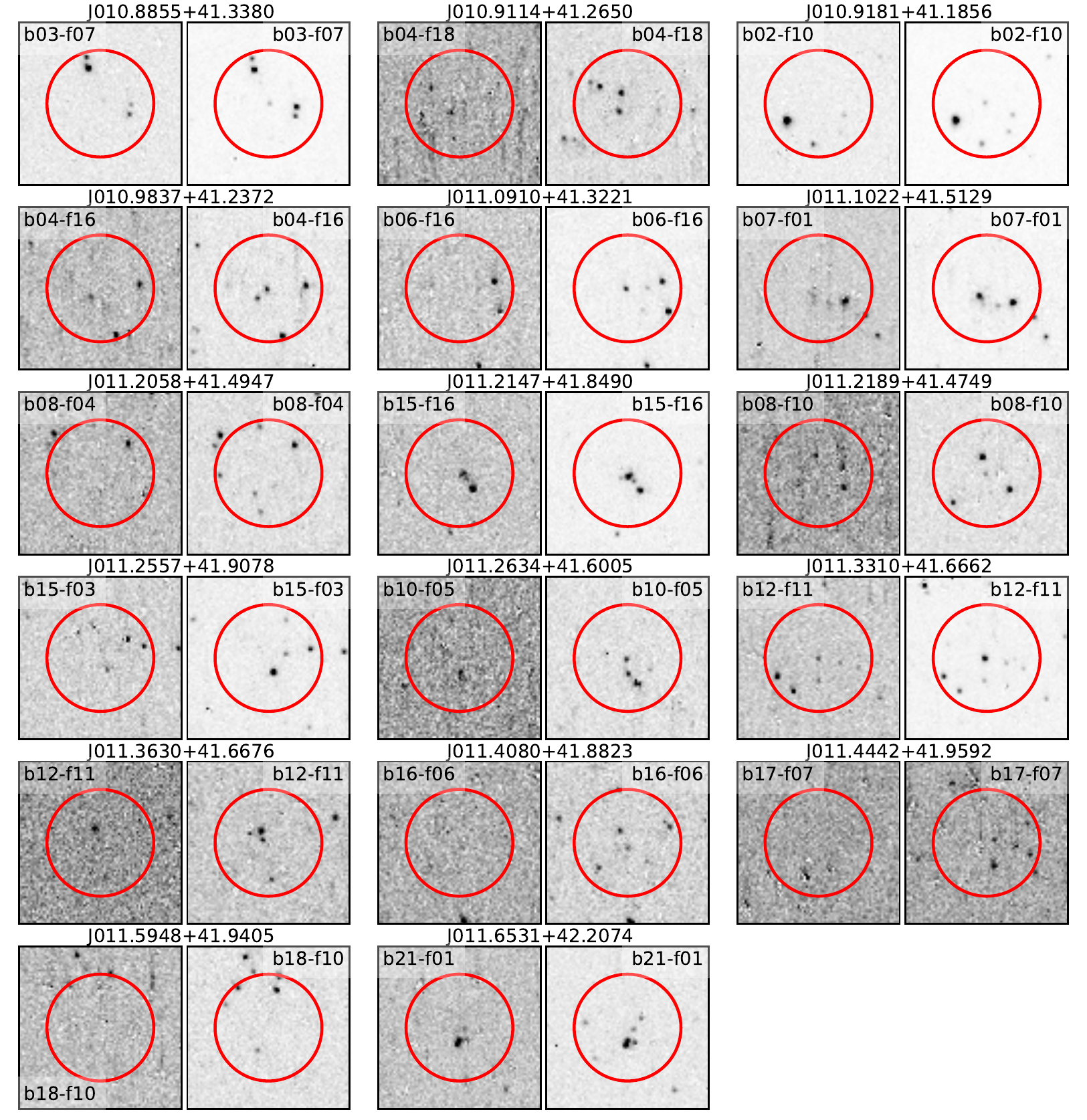}
\caption{Image cutouts near PS1 Cepheids with possible overdensities of stars that are not clearly significant clusters, cf. Fig.\,\ref{app:fig:UVPS_CCs}.\label{app:fig:UVPS_interesting}}
\end{figure*}

\begin{figure*}
\centering
\includegraphics{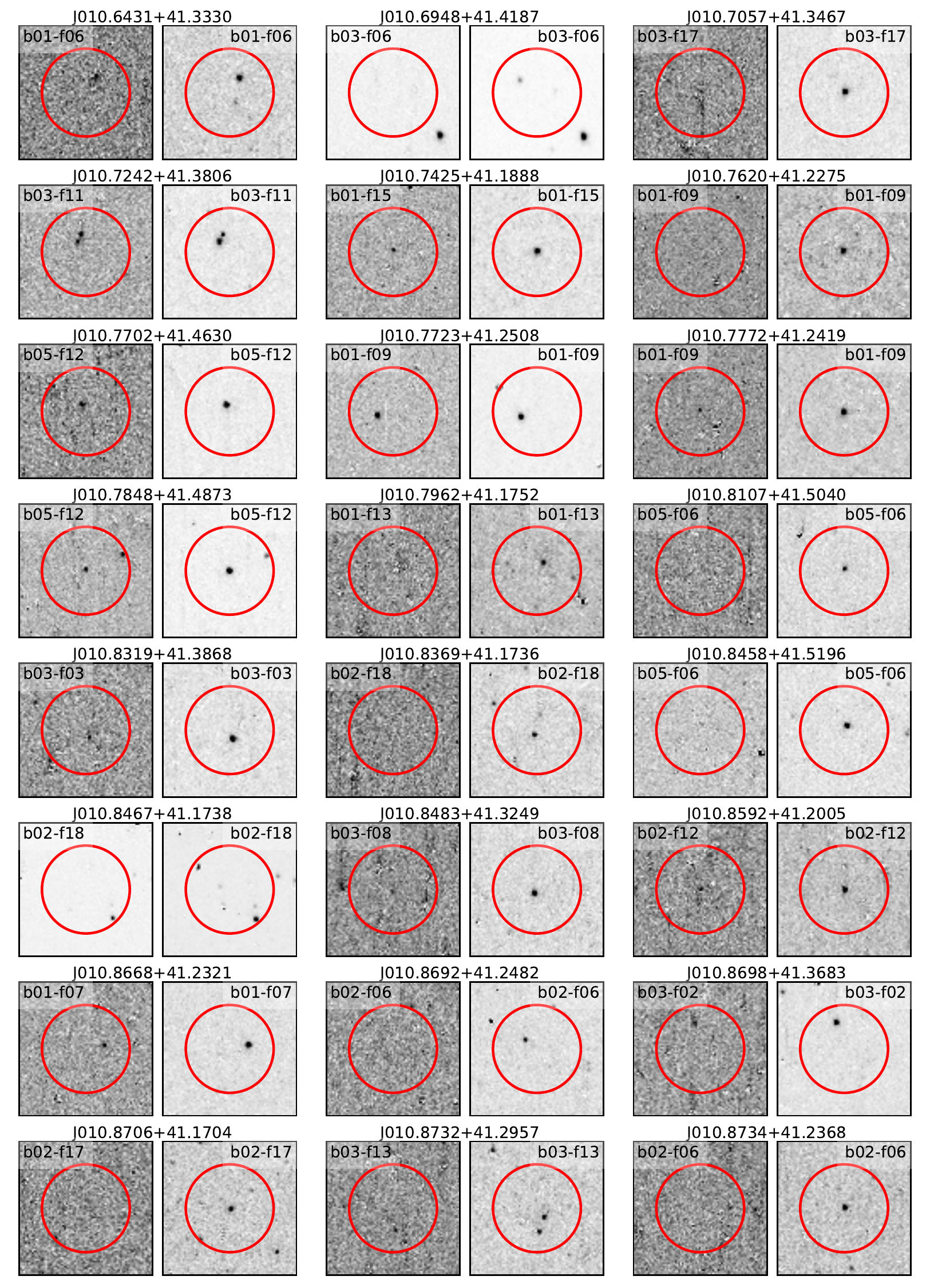}
\end{figure*}

\begin{figure*}
\centering
\includegraphics{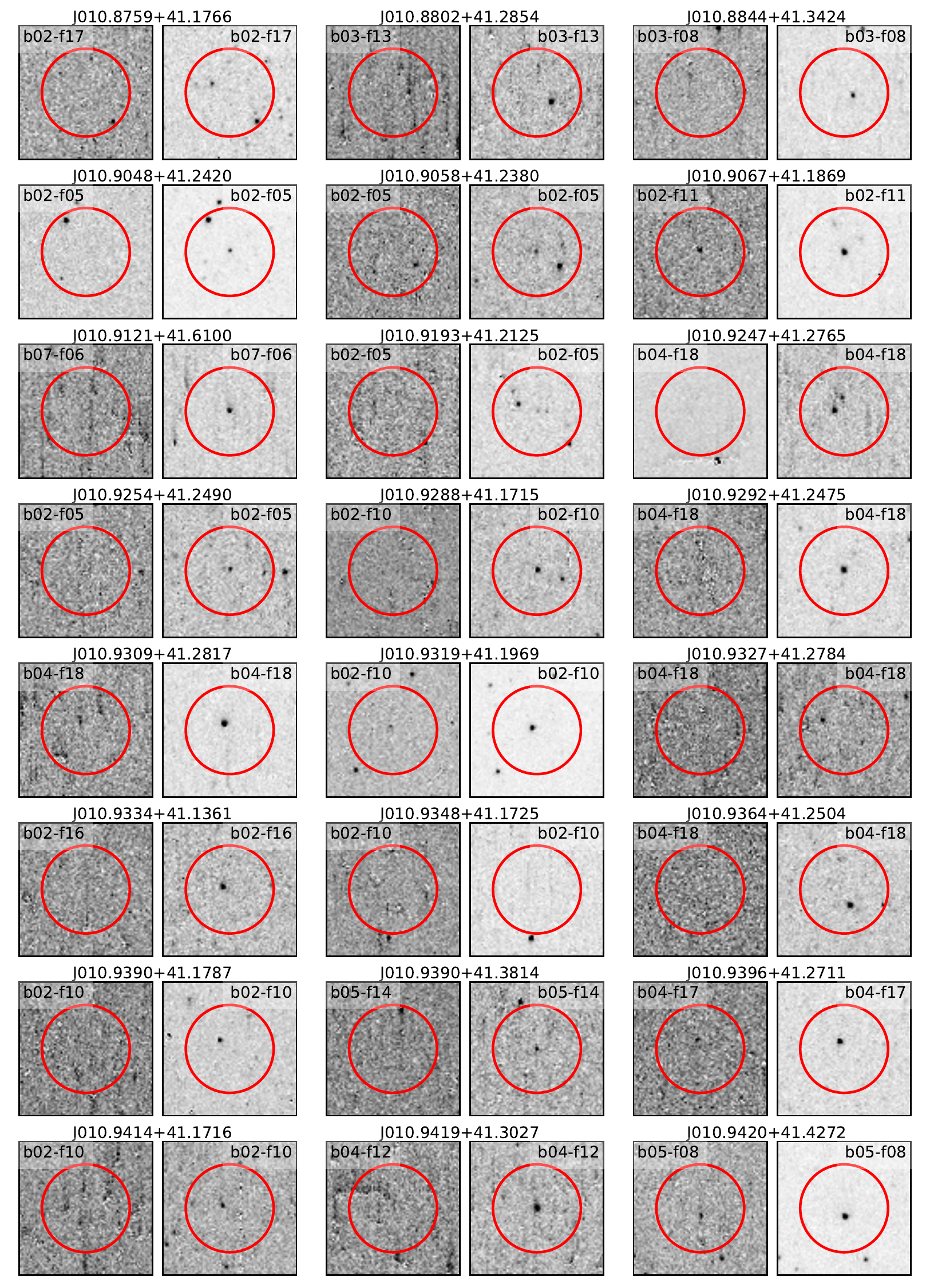}
\end{figure*}

\begin{figure*}
\centering
\includegraphics{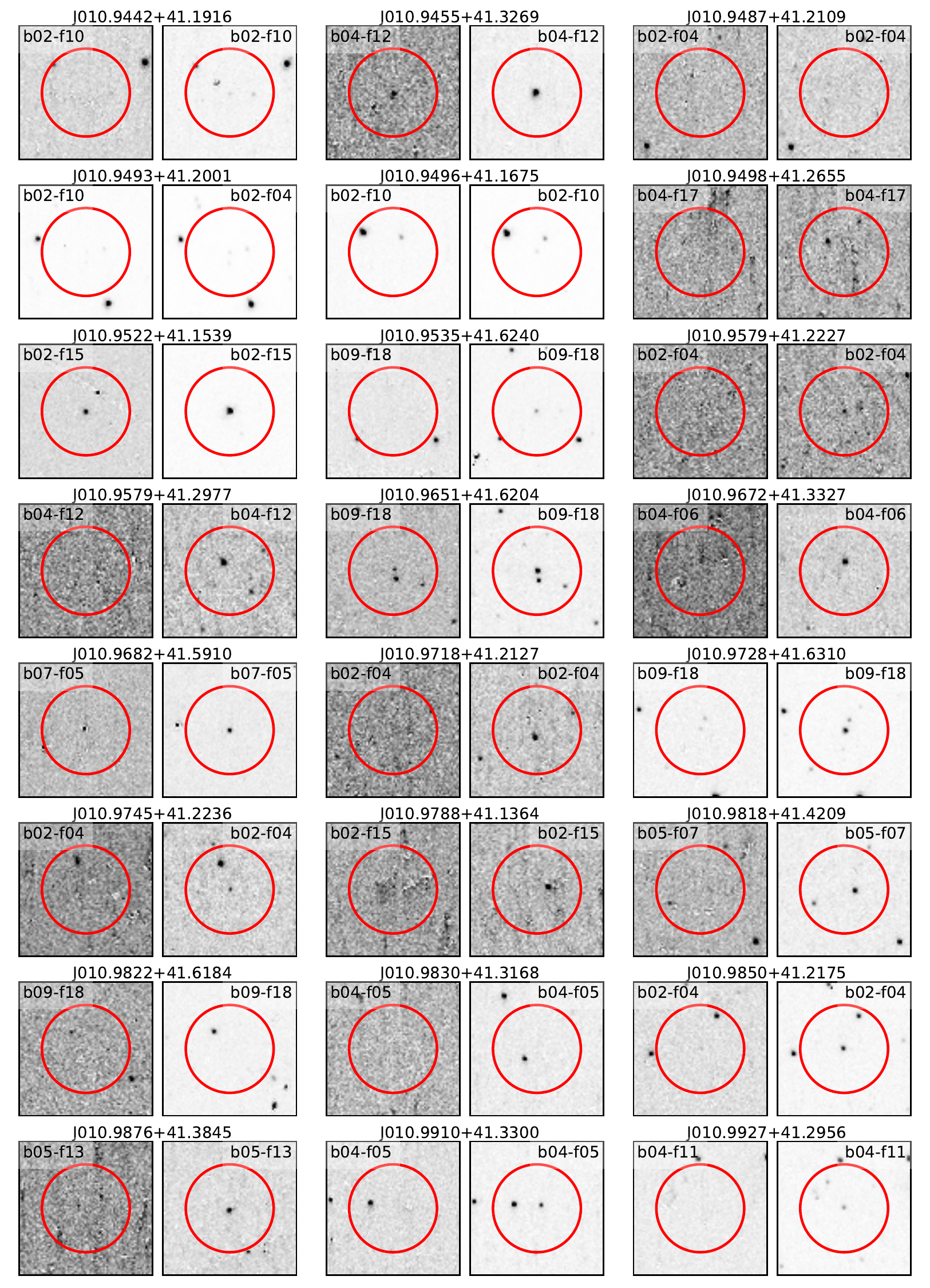}
\end{figure*}

\begin{figure*}
\centering
\includegraphics{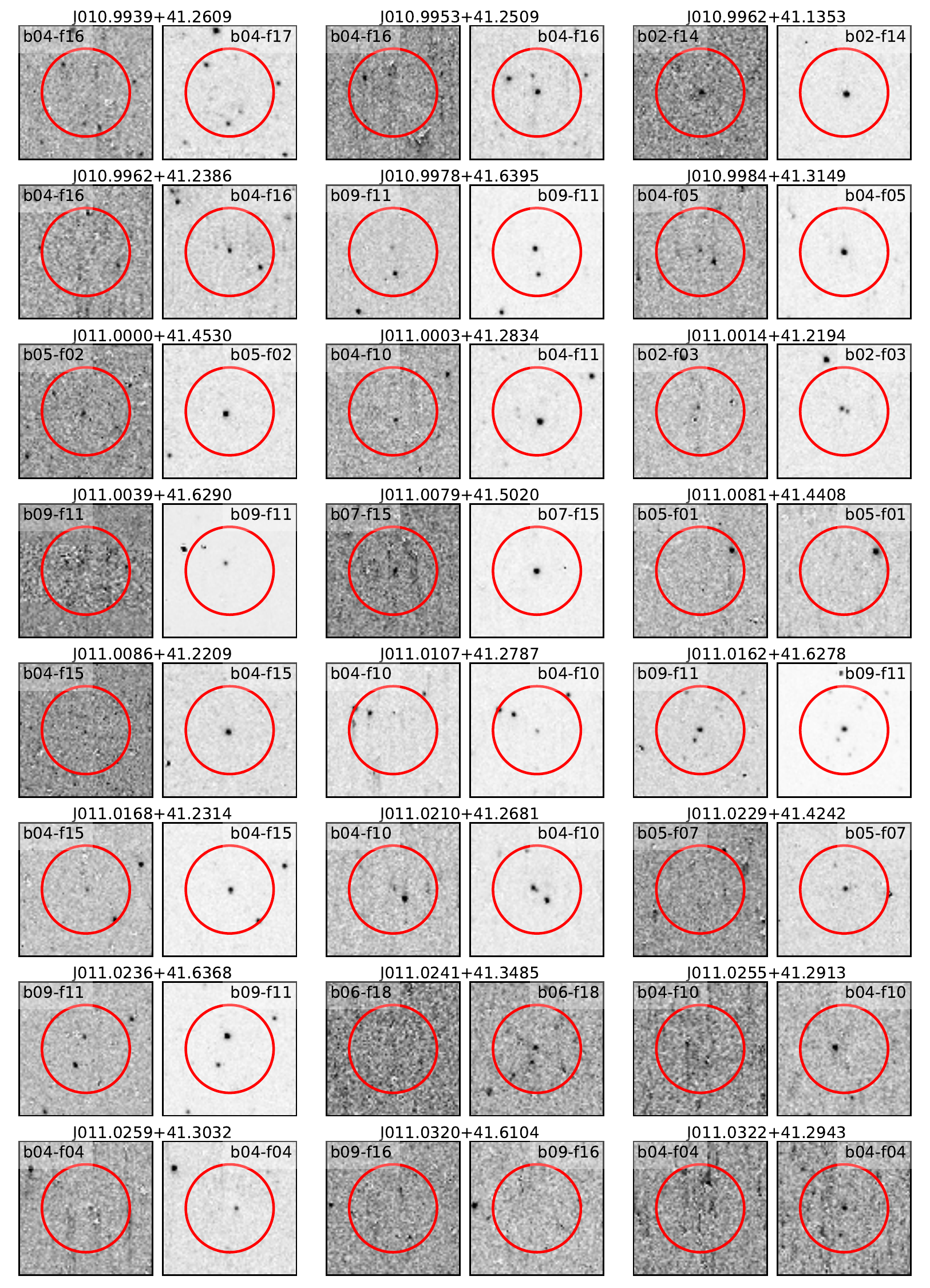}
\end{figure*}

\begin{figure*}
\centering
\includegraphics{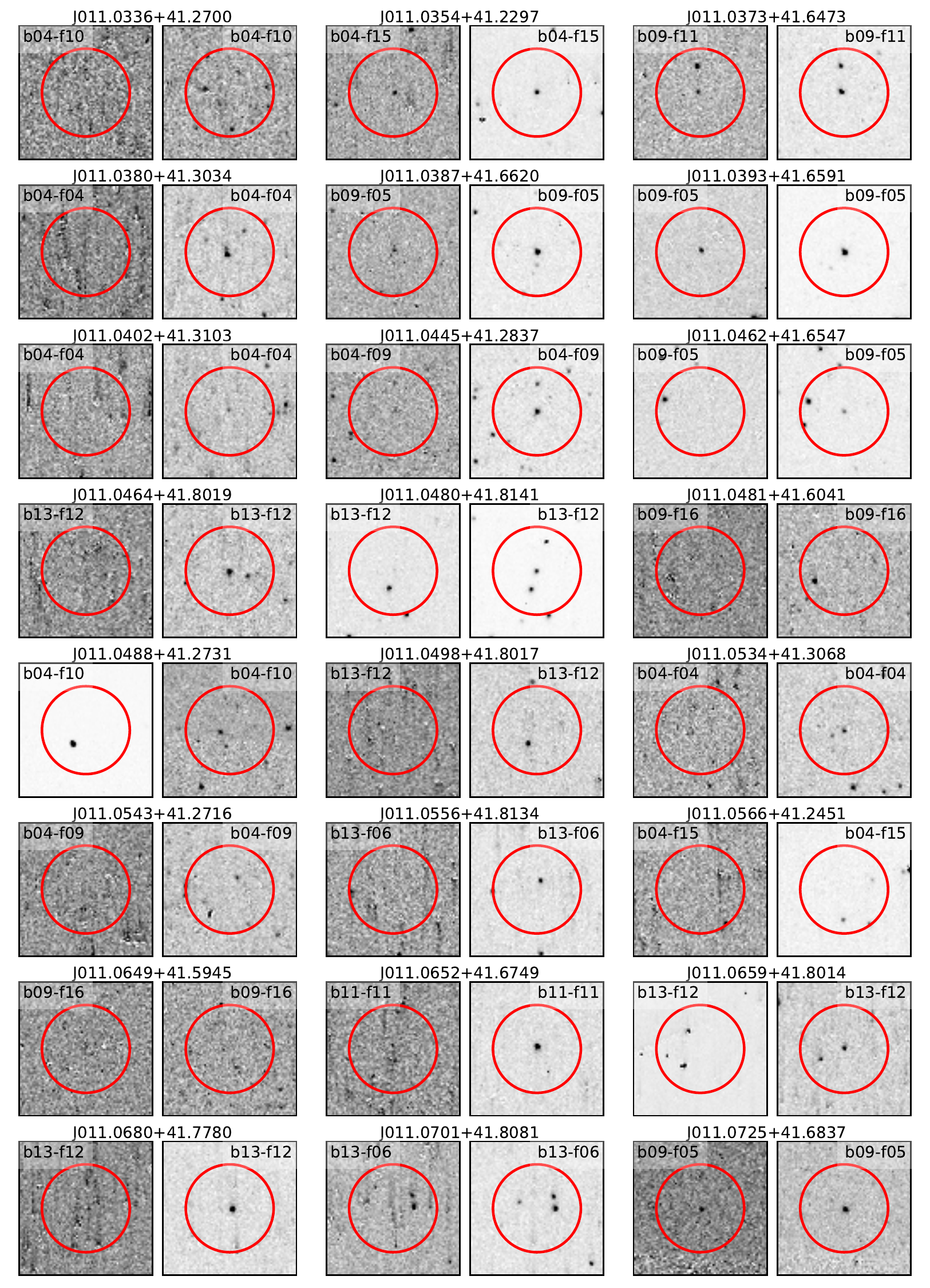}
\end{figure*}

\begin{figure*}
\centering
\includegraphics{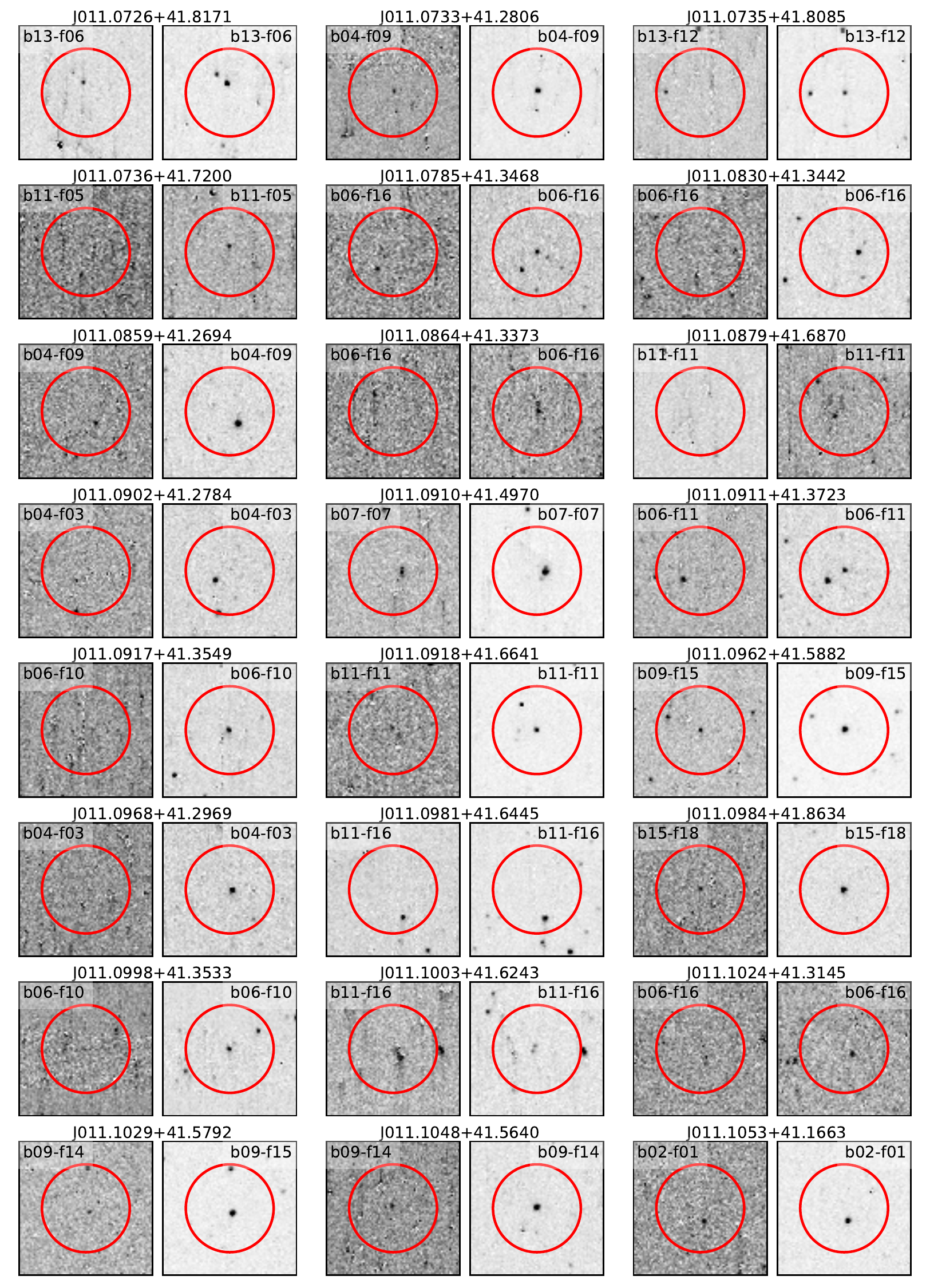}
\end{figure*}

\begin{figure*}
\centering
\includegraphics{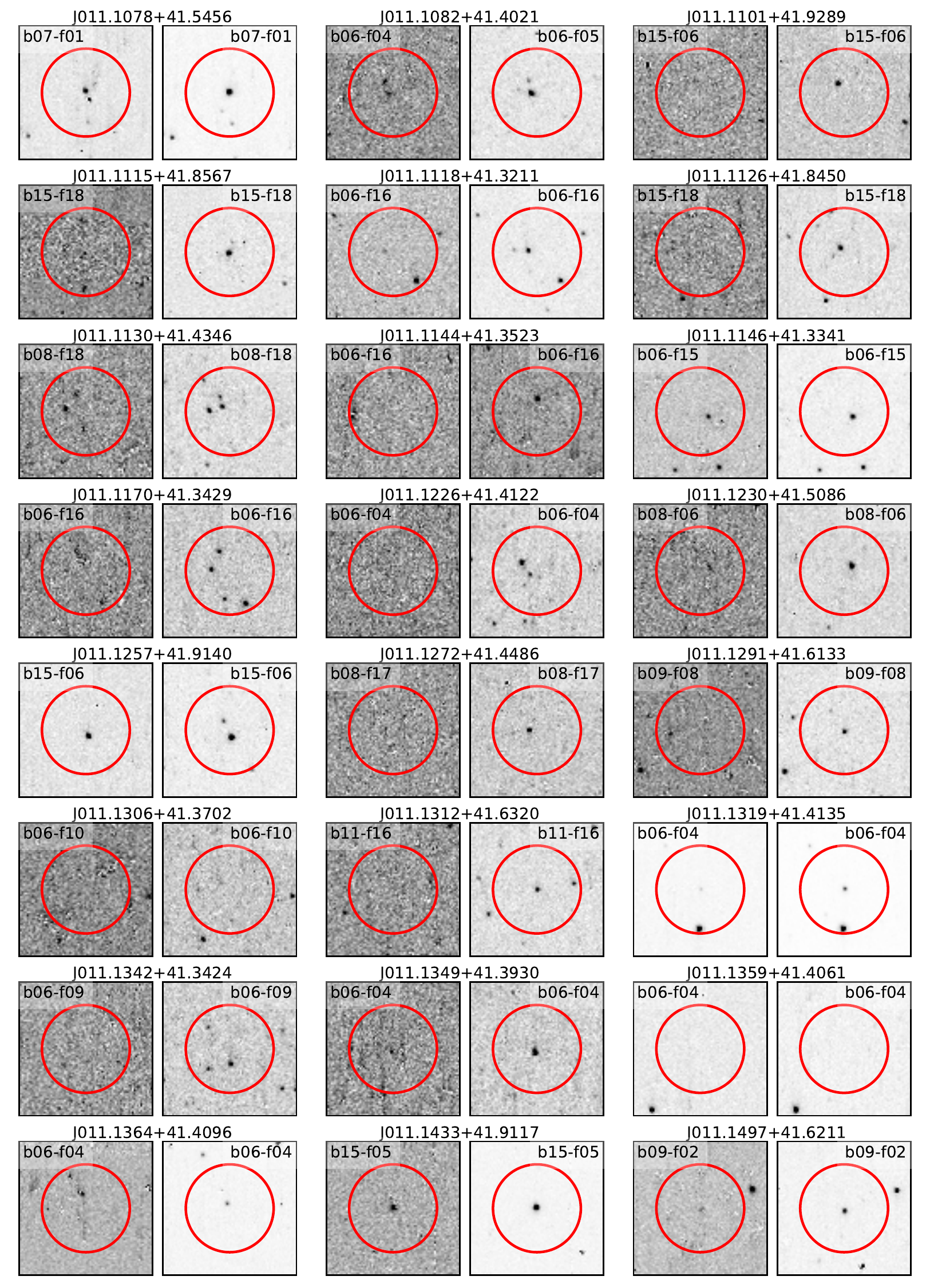}
\end{figure*}

\begin{figure*}
\centering
\includegraphics{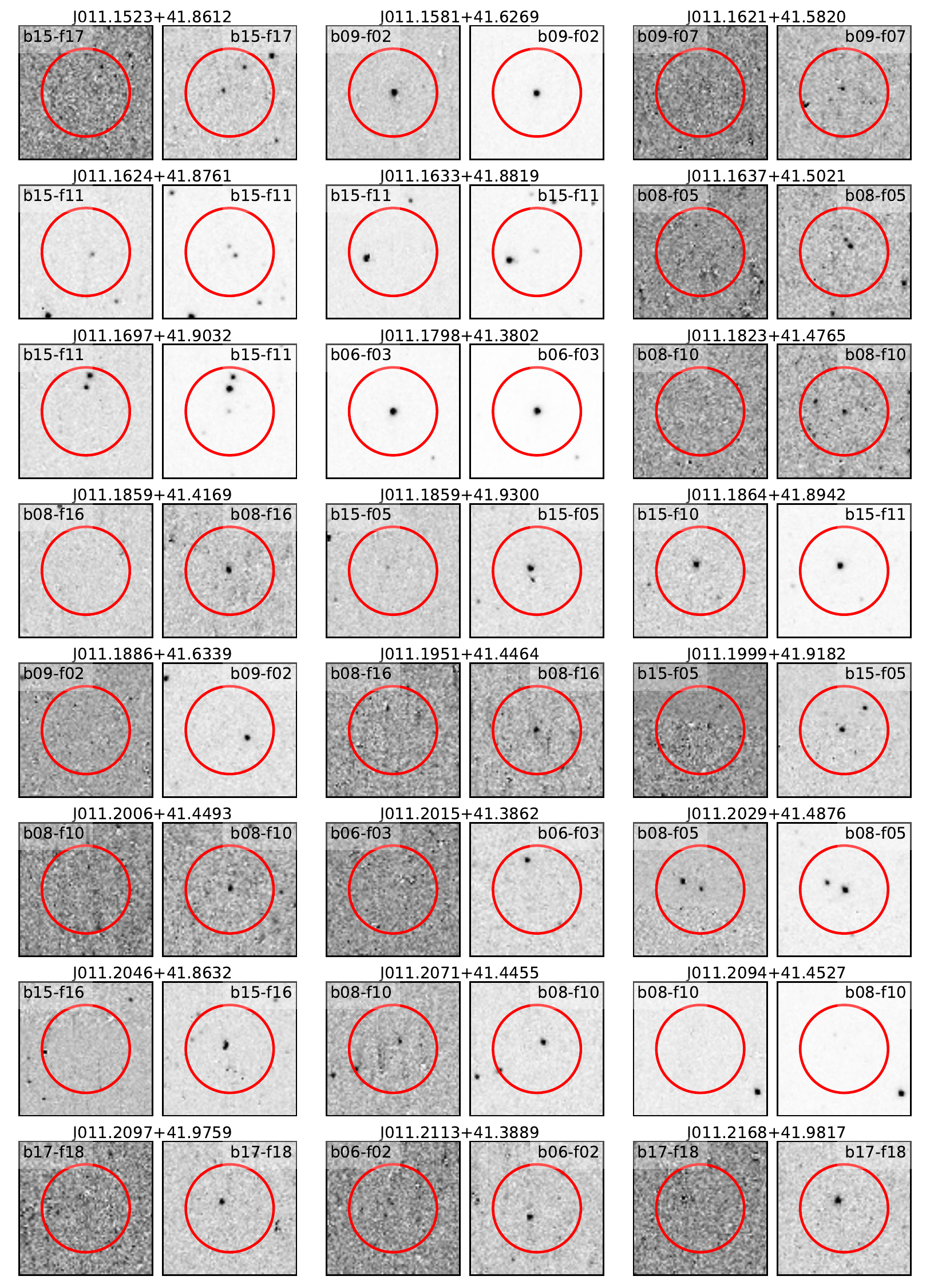}
\end{figure*}

\begin{figure*}
\centering
\includegraphics{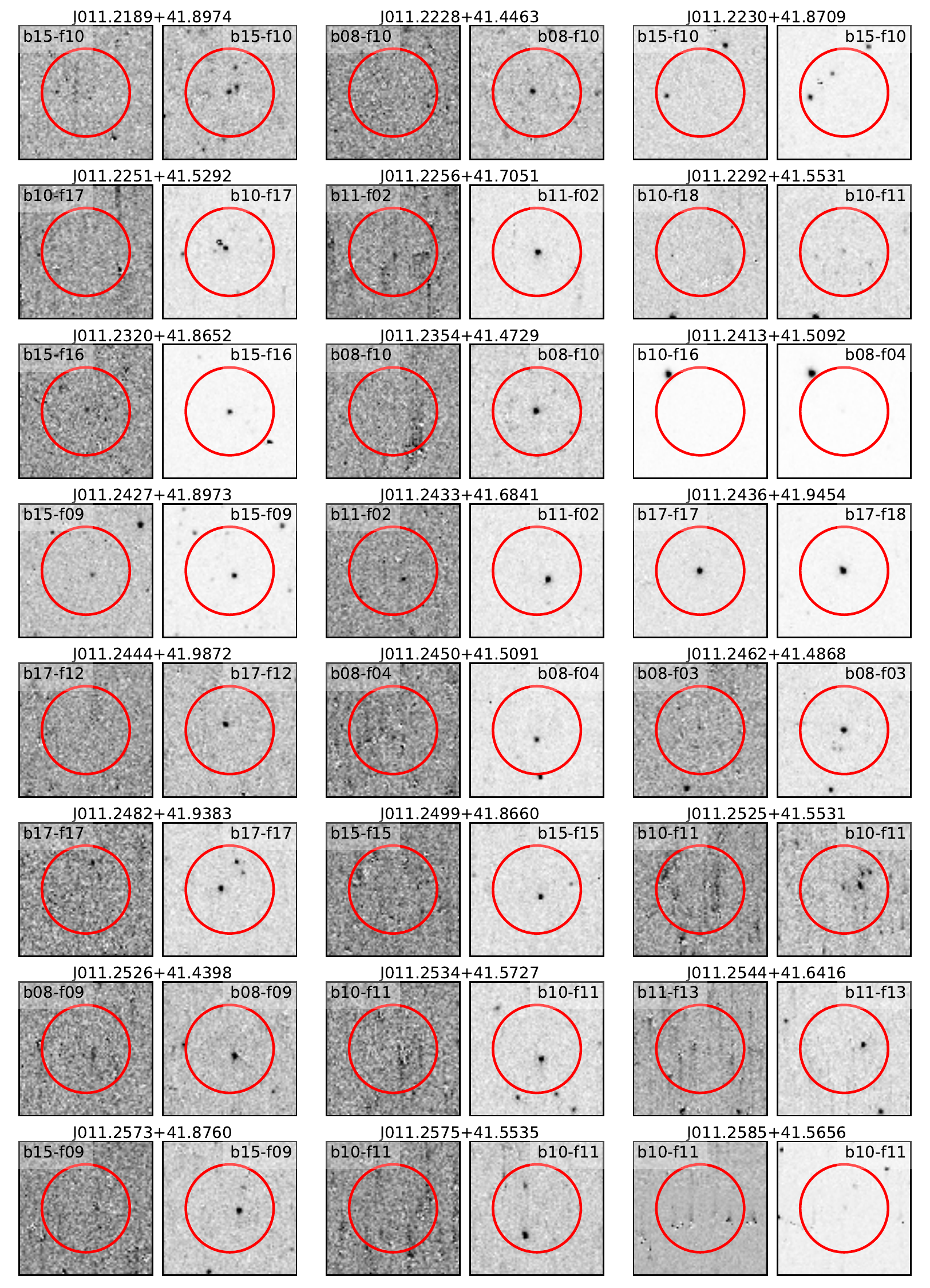}
\end{figure*}

\begin{figure*}
\centering
\includegraphics{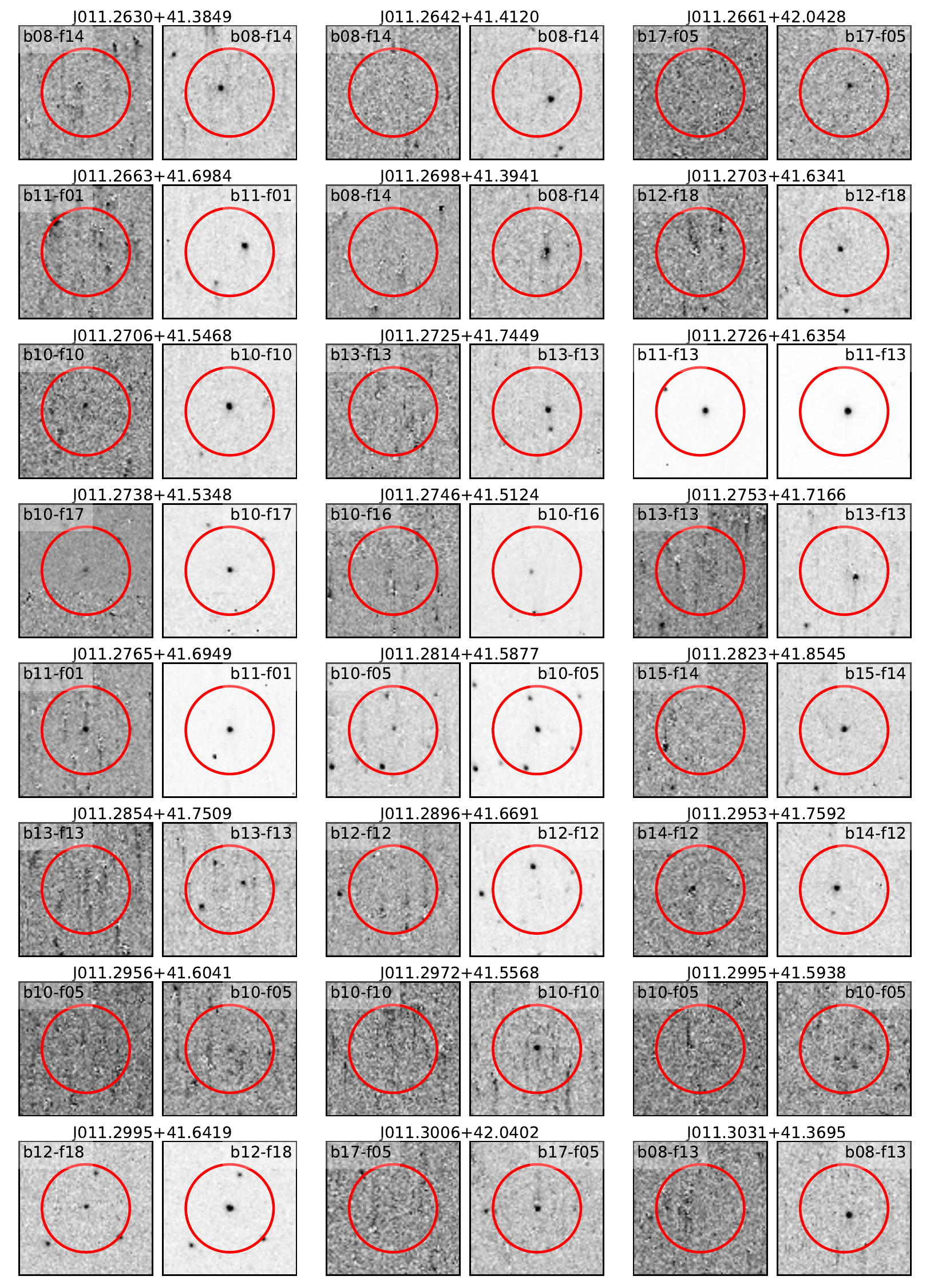}
\end{figure*}

\begin{figure*}
\centering
\includegraphics{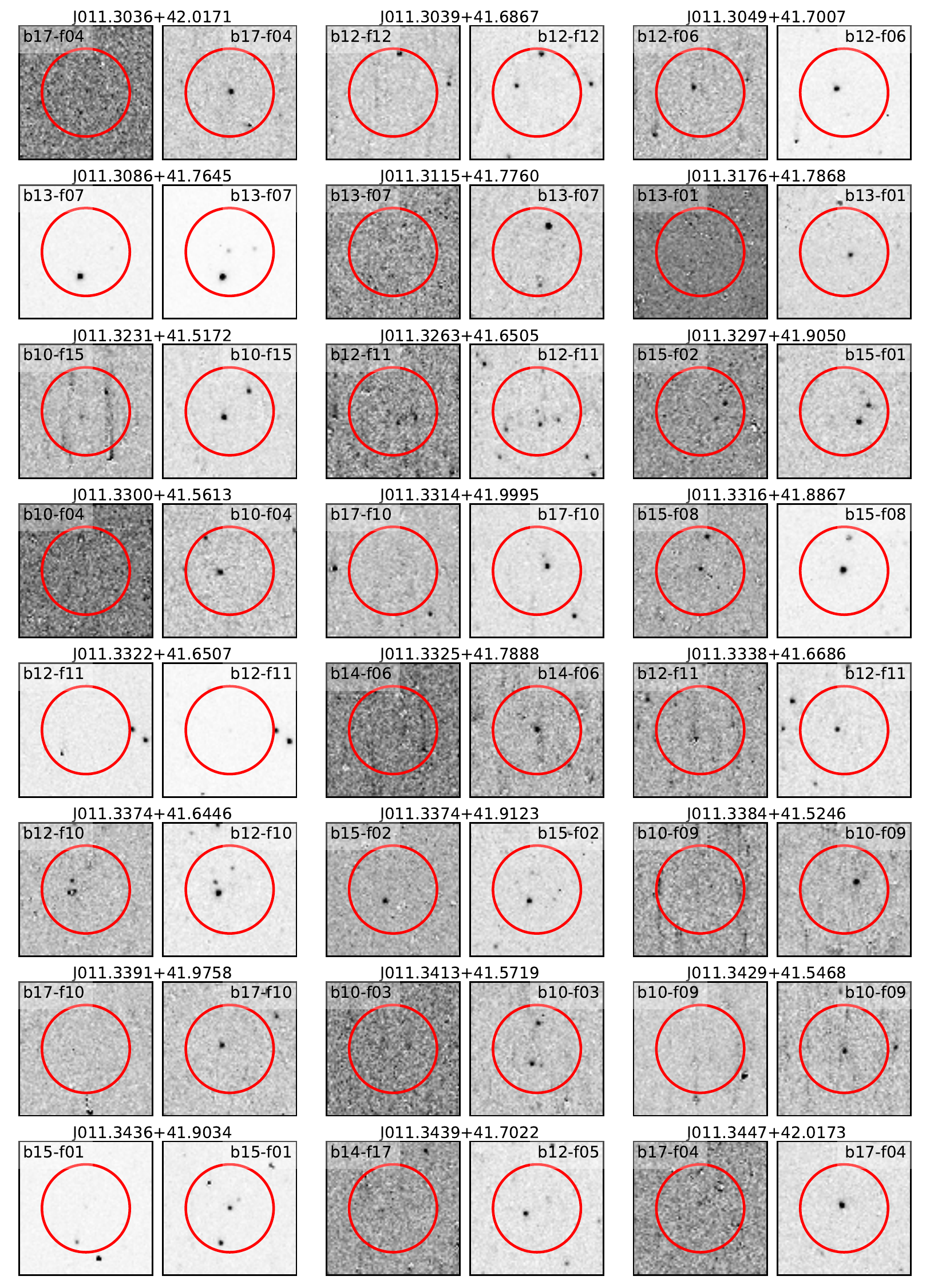}
\end{figure*}

\begin{figure*}
\centering
\includegraphics{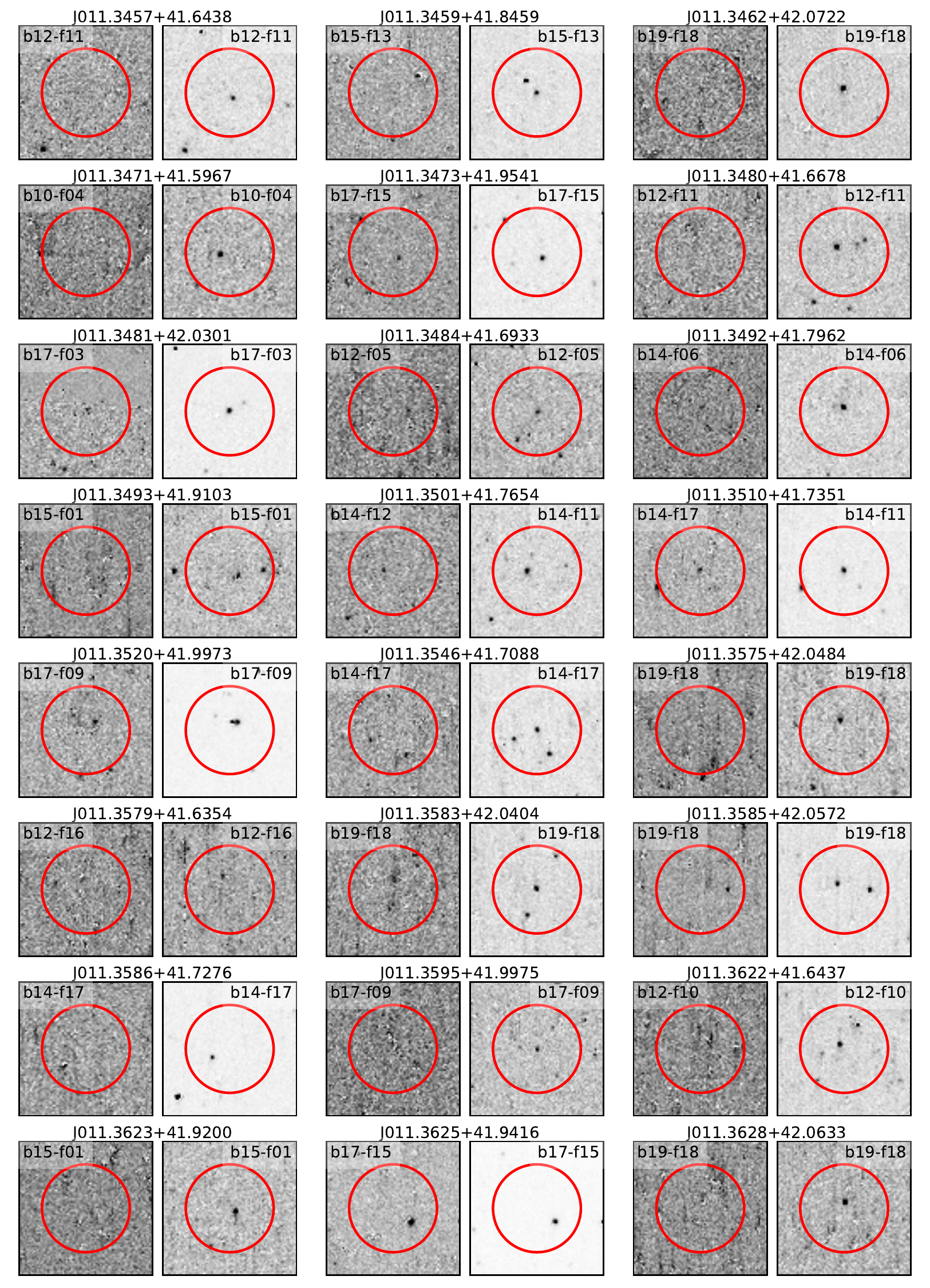}
\end{figure*}

\begin{figure*}
\centering
\includegraphics{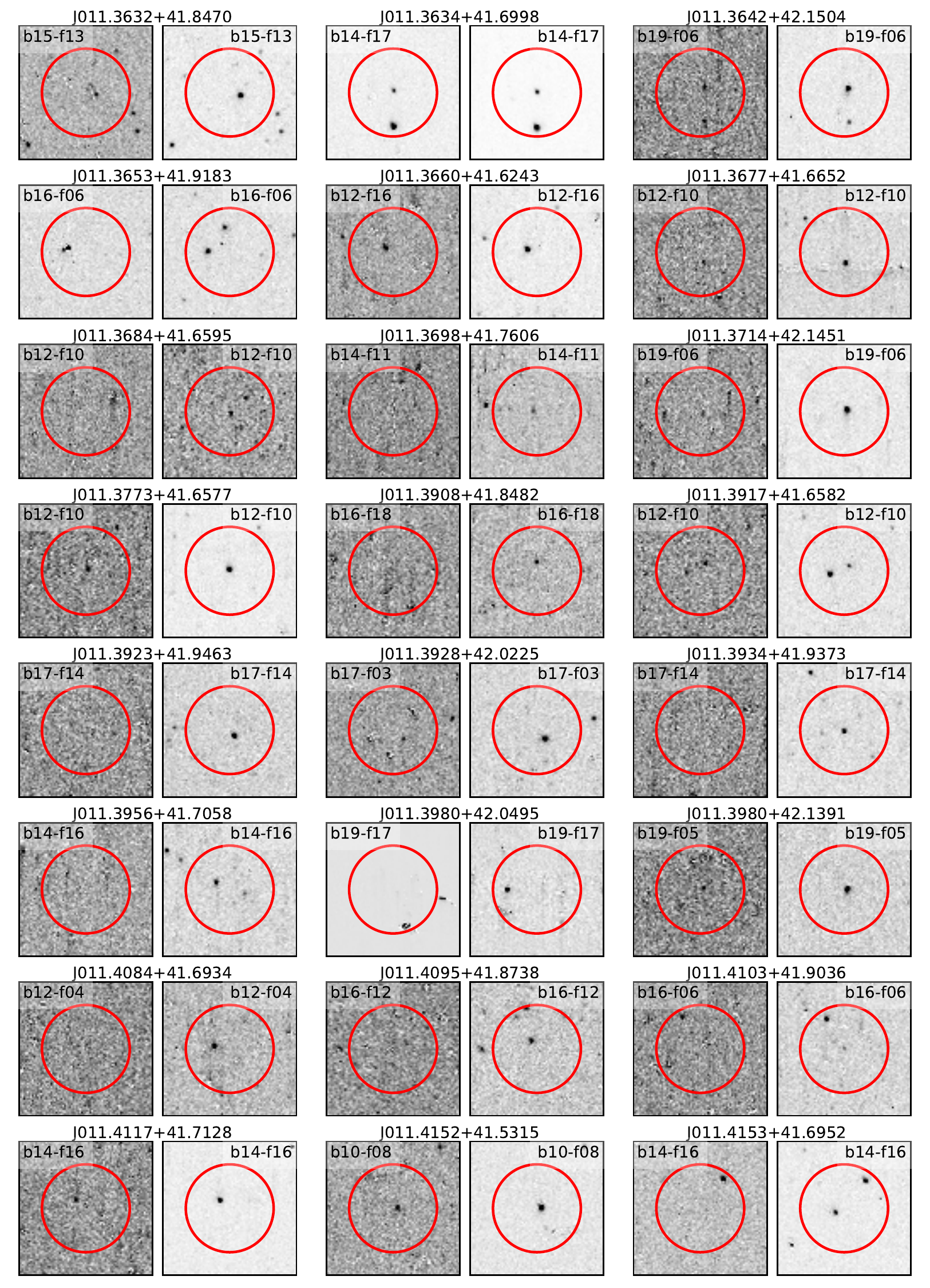}
\end{figure*}

\begin{figure*}
\centering
\includegraphics{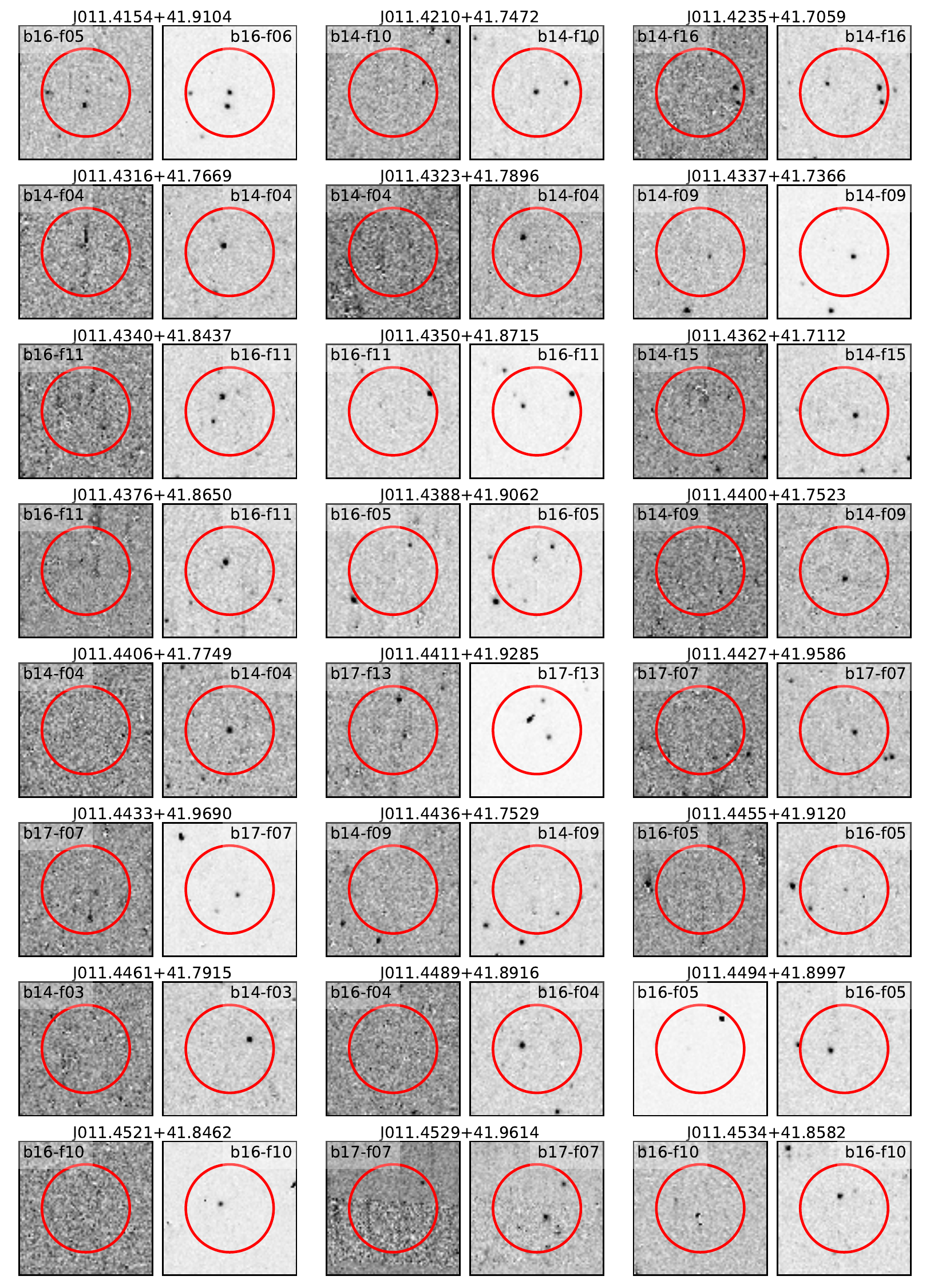}
\end{figure*}

\begin{figure*}
\centering
\includegraphics{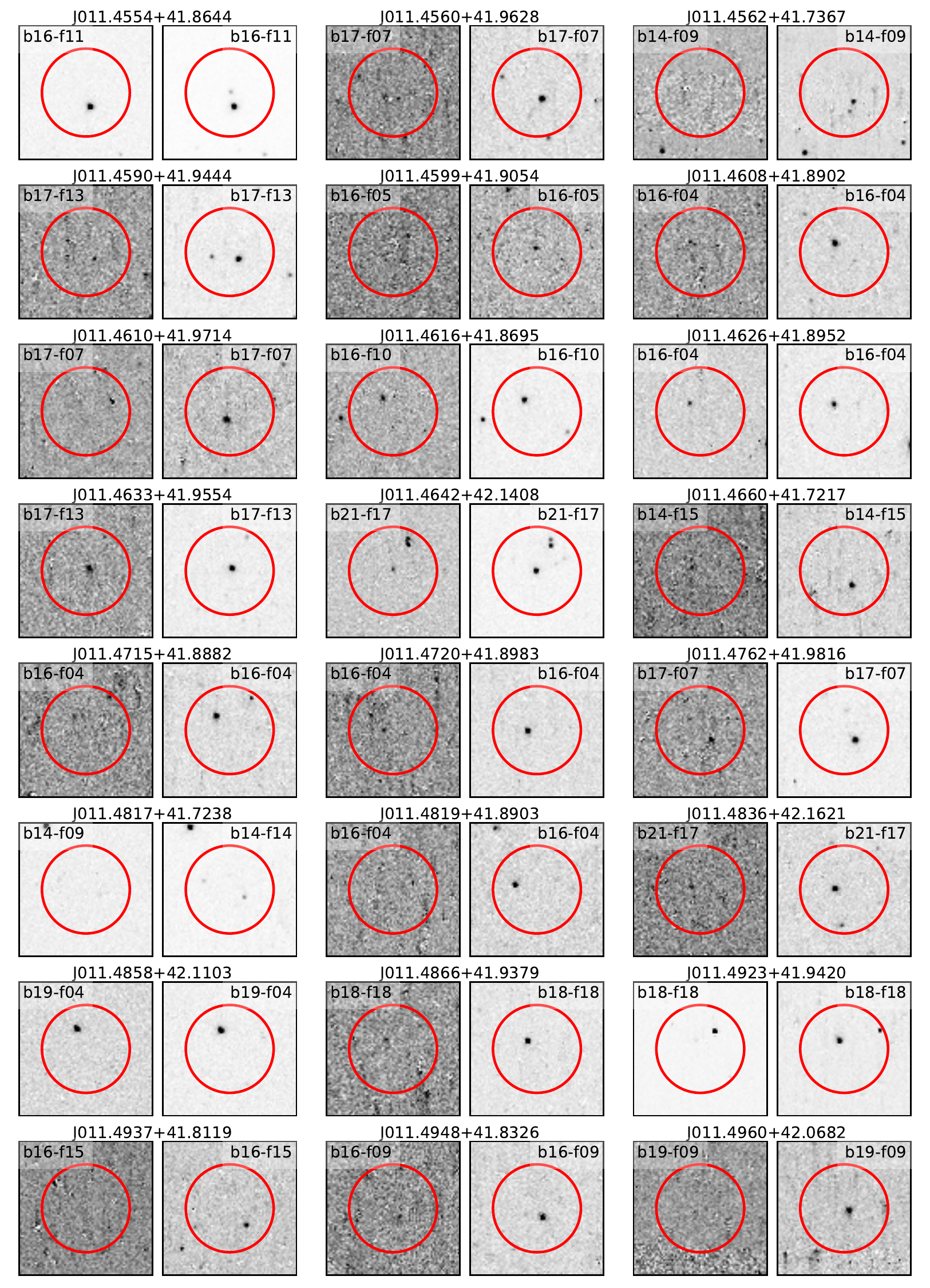}
\end{figure*}

\begin{figure*}
\centering
\includegraphics{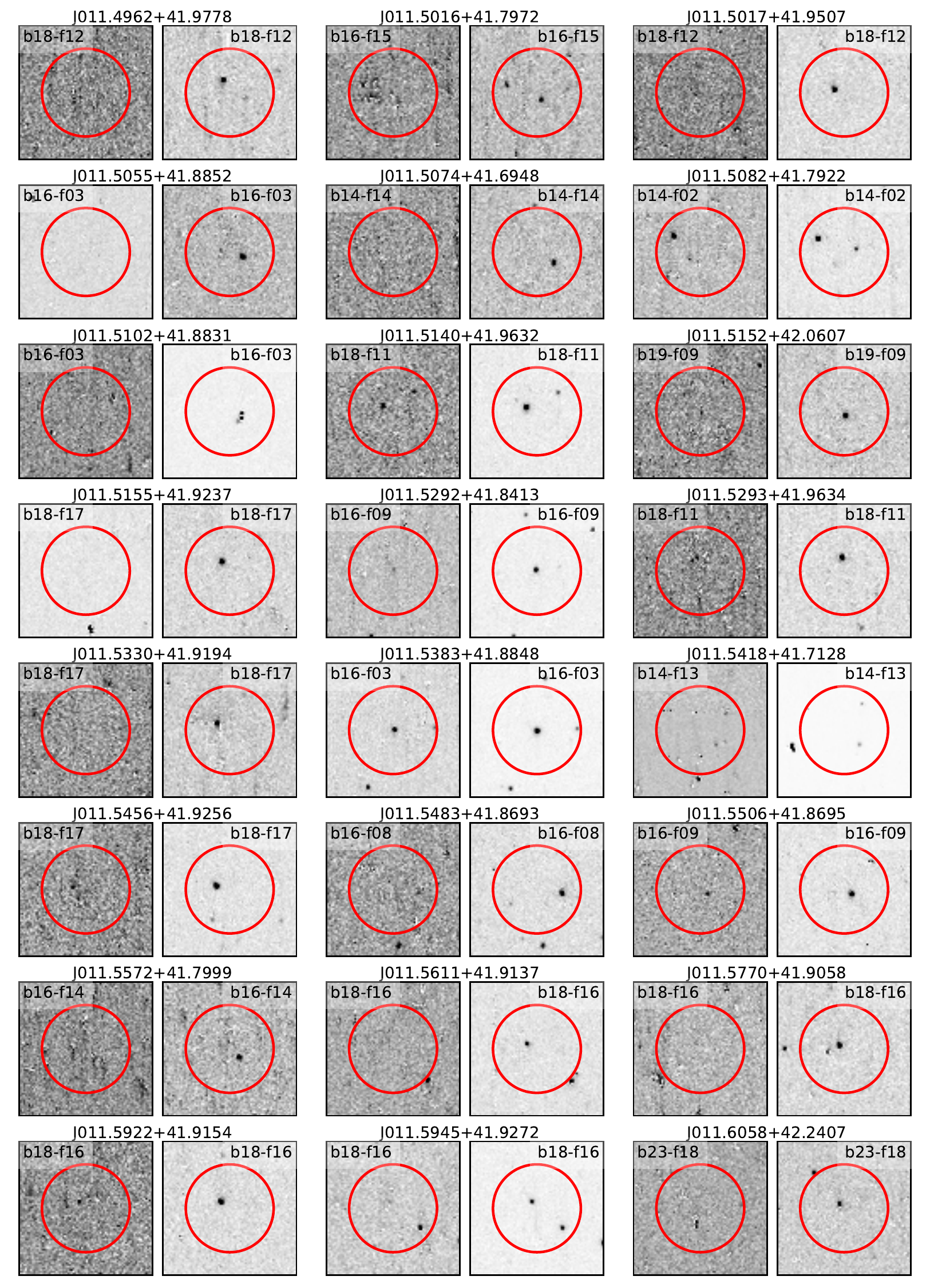}
\end{figure*}

\begin{figure*}
\centering
\includegraphics{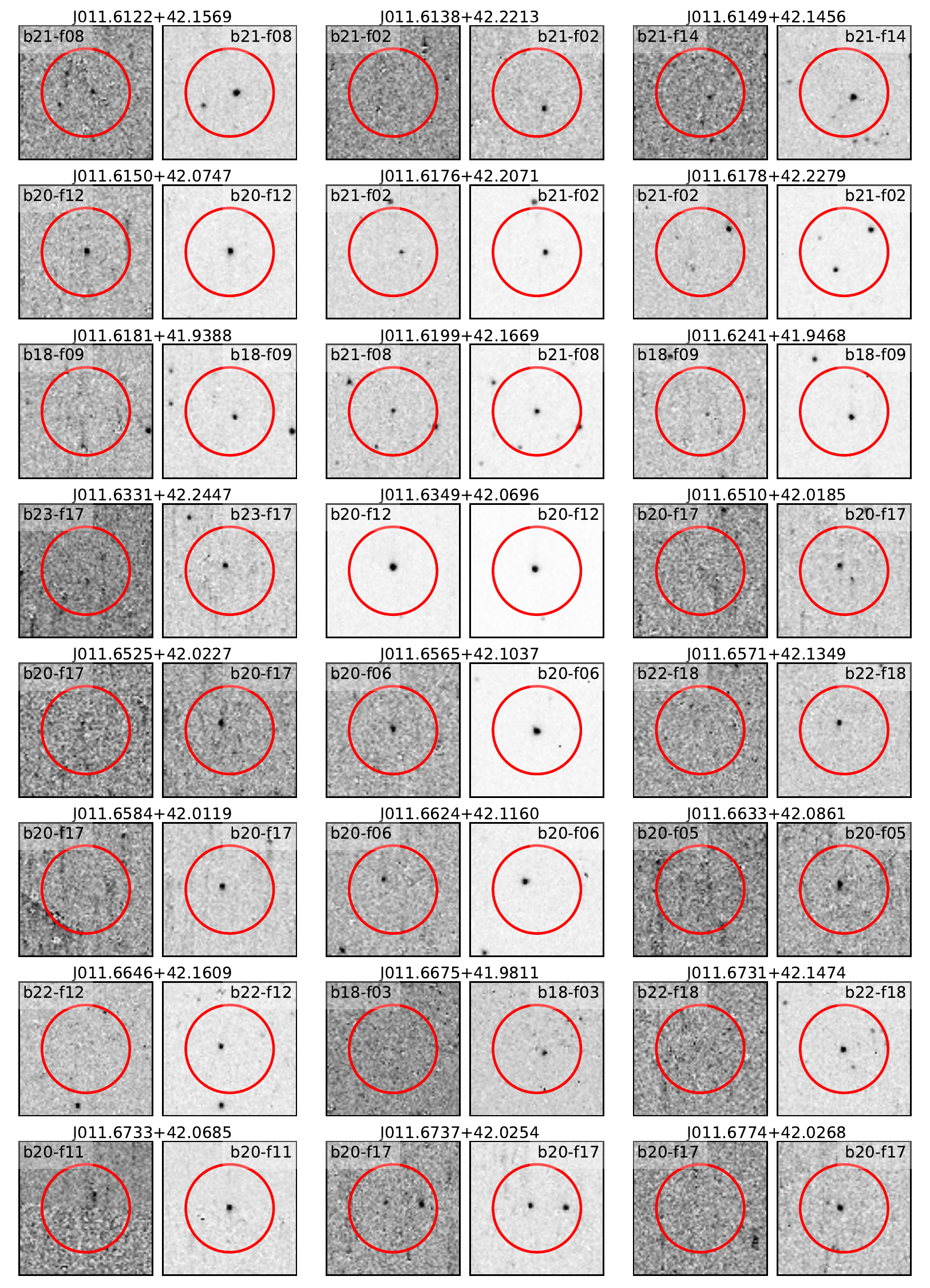}
\end{figure*}

\begin{figure*}
\centering
\includegraphics{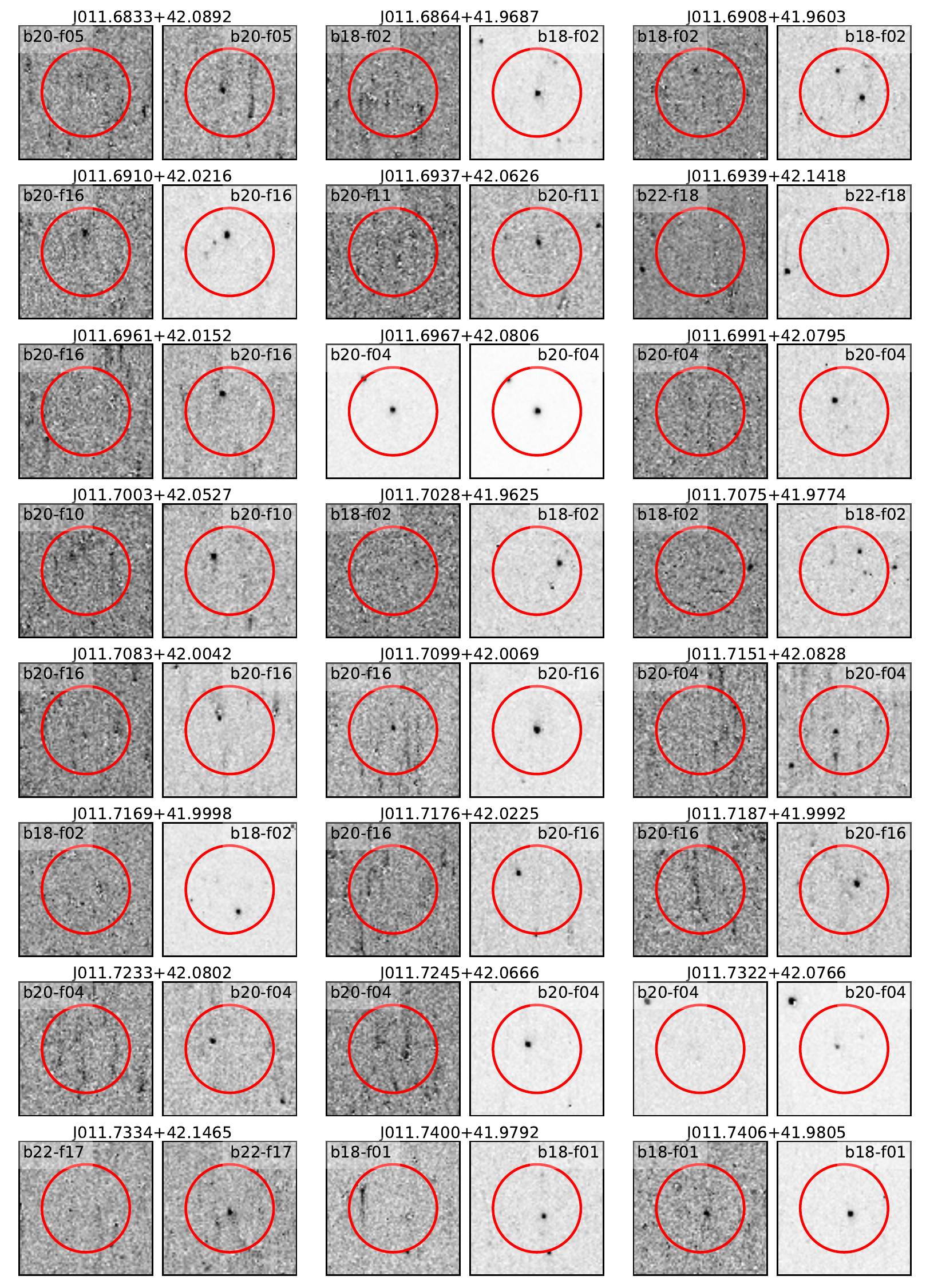}
\end{figure*}

\begin{figure*}
\centering
\includegraphics{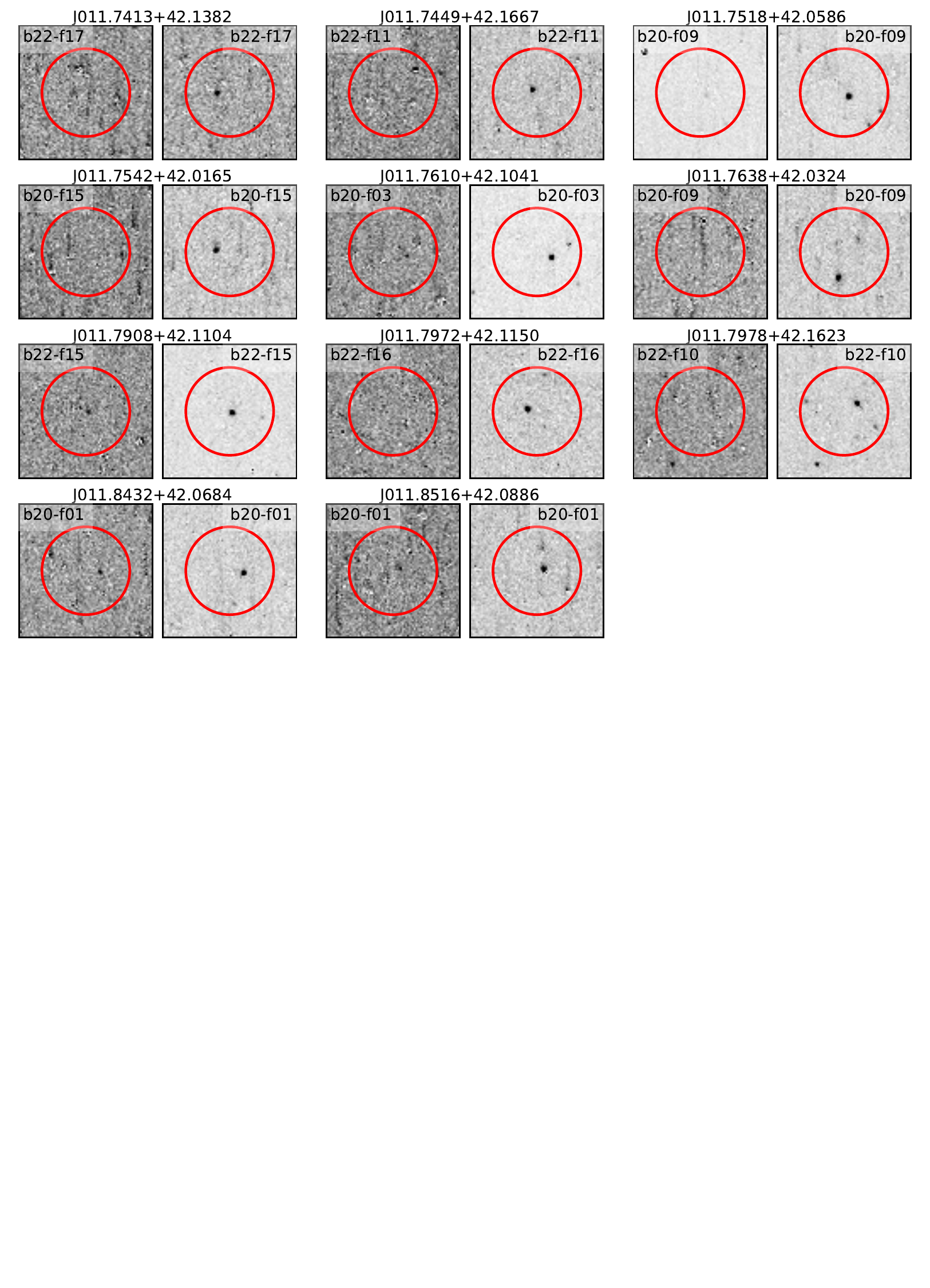}

\caption{UV postage stamps of all remaining fundamental mode (FU) or unclassified (UN) PS1 Cepheids located within the PHAT footprint \citep{2013AJ....145..106K}, see Fig.\,\ref{app:fig:UVPS_CCs} for details.\label{app:fig:UVPS_p18}}
\end{figure*}

\end{appendix}

\end{document}